\documentclass[11pt]{article}

\usepackage{graphicx}
\usepackage{bm}
\usepackage{epsf}
\usepackage{subfigure}
\usepackage{rotating}
\usepackage{epsfig,graphics,rotate,color}
\usepackage{wrapfig}  
\usepackage{amssymb}
\usepackage{amsmath}
\usepackage{amsfonts}
\usepackage{array,hhline,dcolumn}%
   
\advance \topmargin by -\headheight
\advance \topmargin by -\headsep      
     
\evensidemargin \oddsidemargin
\begin{document}

~~~~~\\

\vspace{1.in}
     
~~~~~\\

\begin{center}
{\Large \bf  Cost-effective Design Options for IsoDAR}\\
\vspace{.5 in}
 A.~Adelmann$^1$, J.R.~Alonso$^{2}$, W.~Barletta$^{2}$, R.~Barlow$^{3}$, L.~Bartoszek$^{4}$, A.~Bungau$^3$, L.~Calabretta$^{5}$, A.~Calanna$^{2}$, D.~Campo$^{2}$,
  J.M.~Conrad$^{2}$, Z.~Djurcic$^{6}$, Y.~Kamyshkov$^{7}$, H.~Owen$^{8}$,  M.H.~Shaevitz$^{9}$, I.~Shimizu$^{10}$,
  T.~Smidt$^2$, J.~Spitz$^2$, M.~Toups$^2$,  M.~Wascko$^{11}$, L.A.~Winslow$^{2}$, J.J.~Yang$^{1,2}$

\vspace{.1 in}
{\it  
$^1$Paul Scherrer Institut, Villigen, CH-5232, Switzerland \\
$^2$Massachusetts Institute of Technology, Cambridge, MA 02139, USA \\
$^3$University of Huddersfield, Huddersfield, HD1 3DH, UK \\
$^4$Bartoszek Engineering, Aurora, IL 60506, USA \\
$^5$Istituto Nazionale di Fisica Nucleare, Laboratori Nazionali del Sud, I-95123, Italy \\
$^6$Argonne National Laboratory, Argonne, IL 60439, USA \\
$^7$University of Tennessee, Knoxville, TN 37996, USA \\
$^8$University of Manchester and Cockcroft Institute, Manchester, M13 9PL, UK\\
$^9$Columbia University, New York, NY 10027, USA \\
$^{10}$Tohoku University, Sendai, 980-8578, Japan \\
$^{11}$Imperial College London, London, SW7 2AZ, United Kingdom }\\
  \end{center}
\vspace{0.5in}

\noindent {\bf Abstract:}  This whitepaper reviews design options for the 
IsoDAR electron antineutrino source. IsoDAR is designed to produce $2.6\times 10^{22}$ $\bar \nu_e$ per year with an average energy of 6.4~MeV, using isotope decay-at-rest.   Aspects which must be balanced for cost-effectiveness
include: overall cost; rate and energy distribution of the $\bar \nu_e$ flux and backgrounds; low technical risk;
compactness;  simplicity of underground construction and operation; reliability; value
to future neutrino physics programs; and value to industry.  We
show that the baseline design outlined here is the most cost effective.

\newpage

\tableofcontents

\newpage

\vspace{0.25in}

This whitepaper reviews the design options for a
decay-at-rest $\bar \nu_e$ source~\cite{PRL}.  ``IsoDAR" is a novel, very high-intensity
source that will produce $\bar \nu_e$ with $\langle E \rangle = 6.4$~MeV 
from the beta decay-at-rest of $^8$Li.   When paired with an
existing $\sim$1~kton scintillator-based
detector, such as KamLAND~\cite{kamland}, IsoDAR opens a wide range of 
possible searches for beyond standard
model physics.  Refs.~\cite{PRL, highlight} discuss the
outstanding potential of a search for sterile
neutrino oscillations using the $L/E$ dependence of inverse beta decay
(IBD) events ($\bar \nu_e + p \rightarrow e^+ + n$) within the detector.   
Beyond this, the source is designed to produce more than an order of
magnitude more $\bar \nu_e$-electron scattering events than all past 
$\bar \nu_e$ experiments, opening up new opportunities for searches beyond the Standard Model~\cite{sin2thwreact,Valle,andre}.    Studying the production of
exotic particles which subsequently decay in the detector is also an 
interesting goal, as this may relate to the dark matter problem~\cite{paraphotons}.

IsoDAR-like $\bar \nu_e$ sources have been considered for 
underground physics in the past~\cite{McDonald, Russians, Russ2}.
However, such sources have not led to sufficiently high event rates to
reach physics goals.   In Ref.~\cite{PRL}, we have
presented a design which can provide the physics measurements mentioned above.
This conceptual design arose while considering an injector for the DAE$\delta$ALUS (Decay-At-rest Experiment for $\delta_{CP}$ studies
At a Laboratory for Underground Science) $CP$-violation experiment~\cite{NIM, EOI, firstpaper}.    It was recognized that this injector
cyclotron could be used as an intense source of $\bar \nu_e$. IsoDAR can therefore be considered as part of a phased program to establish the DAE$\delta$ALUS program.

The base IsoDAR design utilizes a 60~MeV/n
high-power compact (as opposed to separated sector) cyclotron.
The cyclotron accelerates 5~mA of H$_2^+$ injected
from a conventional ion source at 70~keV.  This
delivers 10~mA of protons (600~kW, continuous wave) to target as H$_2^+$ is two protons bound by one electron. The beam impinges on a beryllium target,
which produces a very high neutron flux.  The neutrons 
enter a surrounding 99.99\% pure $^7$Li sleeve, thermalize, and capture to produce $^8$Li.  The beta-decay of $^8$Li produces the $\bar \nu_e$ that
can be detected in an adjacent scintillator-based detector like KamLAND.  We assume a vertex resolution of 12~cm and an energy resolution 6.4\%/$\sqrt{E~\mathrm{(MeV)}}$ in the detector, consistent with KamLAND~\cite{kamlandres}.  
The most relevant experimental parameters are listed in Table~\ref{assumptions}.

\begin{table}[t]
\label{base_opti}
  \begin{center}
    {
      \begin{tabular}{|c|c|} \hline
        Accelerator  & 60~MeV/amu of H$_2^+$  \\  \hline
        Current  & 10~mA of protons on target  \\  \hline
        Power  & 600~kW  \\  \hline
        Up-time   & 90\%  \\  \hline
        Run period  & 5~years (4.5~years live time)  \\  \hline
        Target   & $^9$Be surrounded by $^7$Li (99.99\%)  \\  \hline
        $\overline{\nu}$ source  & $^8$Li $\beta$ decay ($\langle
E_\nu\rangle$=6.4~MeV)  \\  \hline
        $\overline{\nu}_e$/1000 protons  &14.6 \\  \hline
        Total flux during run & 1.29$\times10^{23}$
$\overline{\nu}_e$ \\  \hline \hline
        Detector  & KamLAND   \\  \hline
        Fiducial mass  & 897 tons   \\  \hline
        Target face to detector center  & 16~m   \\  \hline
        Reconstruction efficiency  & 92\%   \\  \hline
        Vertex resolution  & 12~cm/$\sqrt{E~\mathrm{(MeV)}}$ \\  \hline
        Energy resolution  & 6.4\%/$\sqrt{E~\mathrm{(MeV)}}$  \\  \hline
        Prompt energy threshold & 3~MeV  \\  \hline
        IBD event total & 8.2$\times 10^5$ \\  \hline
        $\overline{\nu}_e$-electron event total & 7200 \\ \hline
      \end{tabular}
      \caption{The relevant experimental parameters used in this
study, reprinted from Ref.~\cite{PRL}.}\label{assumptions}
}
\end{center}
\end{table}
This base design represents the most cost-effective option for IsoDAR. Here, we present the broad alternatives to this and explain the base design requirements.
We begin in Sec.~\ref{criteria} by discussing the criteria for cost-effectiveness, based on the physics goals mentioned above.  
This is followed by a detailed discussion of our base design in Sec.~\ref{base}.
We explain the partnerships which may significantly add to
cost-effectiveness in Sec.~\ref{widerimplications}, many of which are available because of
the baseline design choices. We then consider alternatives to the base 
design.  These considerations divide into three areas of study:
(1) the fundamental underlying choices of beam particle species,
energy, and target material, discussed in Sec.~\ref{overallalt}; (2) the technology used to
deliver the beam to the target, discussed in Sec.~\ref{driveralt}; and (3) radical
alternatives that are dissimilar to the IsoDAR base design, discussed in Sec.~\ref{radicalalt}.   
In this analysis, we compare to four specific alternatives
which appear to be the only other possible options for this source:
\begin{enumerate}
\item Radio-frequency quadrupole (RFQ) injection with a separated sector cyclotron.
\item A 30~MeV linear accelerator (LINAC) with a 40~mA proton beam.
\item A $\beta$-beam based design. 
\item A new detector located at an existing beamline.
\end{enumerate}
In Sec.~\ref{conclude}, we conclude that the design outlined in 
Ref.~\cite{PRL} is the most cost effective approach among the
alternatives identified.
\section{Criteria for a Cost Effective Design \label{criteria}}

We study alternative design options based on the criteria below for
cost-effectiveness.  In this section, we explain the motivation for
these criteria.  In the concluding section, we provide parameters for ranking alternative
designs. The criteria are:

\begin{enumerate}
\item {\it Cost:}~~Minimizing the total acquisition and operations costs is important to any project.   In the case of running at an existing underground lab, this includes the cost of the accelerator, the target ($\bar \nu_e$ source), electricity, and relevant infrastructure that must be provided to successfully conduct the experiment. When we consider running at an existing accelerator, this will include issues of installing a new beamline and detector with sufficient shielding. 
\item  {\it Rate and energy distribution of the $\bar \nu_e$ flux:}~~Maximizing the $\bar \nu_e$ flux is vital for sensitivity to sterile neutrinos as well as accomplishing the other physics goals.  The IsoDAR design produces
  $2.6\times 10^{22}$ $\bar \nu_e$ per year with a mean energy of 6.4~MeV.   An alternative design must match this or come close.  
  Note that a  mean $\bar \nu_e$ energy well above 3~MeV is important, as this requirement helps differentiate signal events 
  from (usually less energetic) radiogenic backgrounds in the detector.  This is especially
  important for the $\bar \nu_e$-electron scattering and dark matter studies, which rely on single-pulse signals. Note that multiple $\bar \nu_e$ sources with endpoints  $>3$~MeV are acceptable, since the flux is directly
  measured using the IBD interaction and its well-known
  cross section~\cite{PRL,ibdxsec}.
\item  {\it Rate and energy distribution of backgrounds:}~~The 
$\nu_e$ intrinsic background must be minimized with respect to the 
$\bar \nu_e$ flux, and limited to $<3$~MeV. 
  In the case where we
  study the possibility of running at existing accelerators, we also consider cosmogenic
  backgrounds under this item.  This is relevant at low depth since $^8$Li has a half-life
  of 841~ms. A beam spill structure of a few microsecond
  or less, as is offered at many existing
  laboratories, cannot help to distinguish between signal and
  cosmogenic background. 
\item{\it Low technical risk:}~~No existing cyclotron can meet our physics goals.
  Therefore, some  R\&D is required for the base design.  We compare the risk
  involved to the risk of implementing alternative designs.
\item {\it Compactness of both 
accelerator and $\bar \nu_e$ source:}~~Compactness of the accelerator is driven 
by space considerations underground.   The goal is to keep the spatial
footprint of the accelerator within a few meters in all dimensions.
Compactness of the $\bar \nu_e$ source is also important for the
sterile neutrino search, in order to make implementation practical. For a $\bar \nu_e$ event
energy of $\langle E \rangle= 8$~MeV, the sterile neutrino search is
optimized for an oscillation length of about 8~m. Thus, the source, including all of
its shielding, must be located
within about 16~m of the detector in order to successfully reconstruct the oscillation wave.

\item {\it Simplicity of underground construction and operation:}~~Along with compactness, application underground encourages the simplest
possible design, construction, and operation plans.  Cost
considerations here also encourage simplicity.  

\item{\it Reliability:}~~The up-time of the accelerator must
  comfortably exceed the up-time planned for the experiment so as to
 ensure successful completion of the physics goals.     The experiment only requires 10\% downtime in order to measure beam-off backgrounds.    Thus, we seek an
  accelerator which is likely to require low downtime
  for maintenance.

\item {\it Value
to future physics programs:}~~Developing engineering and
infrastructure for future physics programs is desirable.

\item {\it Value of this development to industry:}~~Developing a
  design which is of interest to industry will lead to strong industrial partnerships and sharing of development costs.

\end{enumerate}

We will touch on each of these points throughout the following text.   In the concluding section (Sec.~\ref{conclude}), we return to this itemized discussion.

\section{The IsoDAR Base Design \label{base}}

The major components of IsoDAR are the ion source,
the cyclotron, and the $\bar \nu_e$ source.  These are connected by beam transport systems, the design of which can be considered straightforward.

\subsection{The Ion Source}

\begin{figure}[t]
\begin{center}
{\includegraphics[angle=-0, width=.75\linewidth]{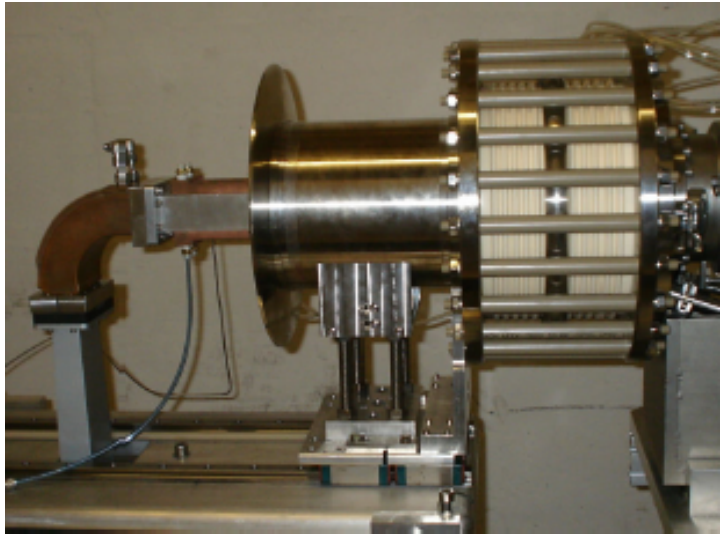}}
\end{center}
\vspace{-0.5in}
\caption{The Versatile Ion Source.
\label{VIS} }
\end{figure}

INFN-Catania has constructed the  Versatile Ion Source (VIS)
(Fig.~\ref{VIS})
\cite{VIS},  which is a non-resonant 
microwave (2.45 GHz) source capable of very high continuous wave (CW) proton or H$_2^+$ beams suitable for IsoDAR. Optimization of one or another ion species is obtained by varying gas pressure and microwave power into the source. The performance of the VIS for production of high-quality,
high-current H$_2^+$ ions will be fully characterized at the BEST Cyclotron Systems test stand,
where the beam will be injected and accelerated to
1~MeV, producing results valuable to 
simulations for IsoDAR and the DAE$\delta$ALUS Superconducting Ring Cyclotron (SRC) beam. 
However, in comparisons to other technology options, 
we regard the overall risk concerning the ion source to be low
since we have already performed preliminary measurements.

The use of H$_2^+$ rather than H$^-$ ions or
protons is a key step forward in development.  
H$_2^+$ is chosen to reduce space charge effects, since it has an 
electric charge of $+1$ for every two protons accelerated. Thus, 5 mA
of this ion corresponds to 10~mA of protons on target. This is discussed further below.

In our design,  a 5~mA
H$_2^+$ beam is injected into the cyclotron at 70~keV (35~keV/amu) 
via a spiral inflector. For comparison,  the generalized
perveance (which parametrizes the
strength of the space charge effect),
\begin{equation}
K=(qI)/(2\pi\epsilon_0 m
\gamma^3 \beta^3)~,  \label{perveance}
\end{equation}
is similar to
that of existing cyclotrons that inject 2~mA of protons at 30~keV~\cite{FRM2CCO04}. This gives us confidence that space-charge forces for the IsoDAR 5~mA H$_2^+$ beam should be manageable. We note, however, that the spiral inflector
must be larger than in these lower-energy proton machines.

\subsection{The Cyclotron \label{isobase}}

\begin{figure}[t]
\begin{center}
\vspace{-0.5in}
{\includegraphics[angle=-0, width=1.\linewidth]{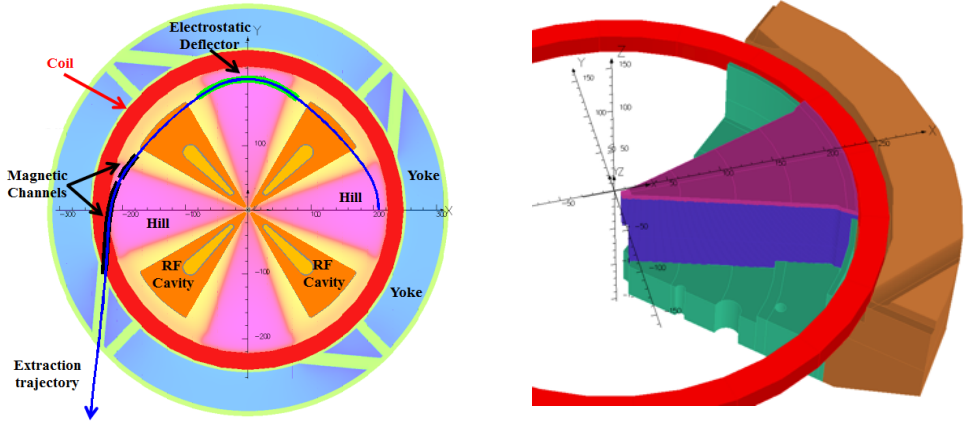}}
\end{center}
\vspace{-2.25in}
\caption{  Left:  Layout of the 
  cyclotron. Pastel colors represent the magnetic field map (magenta is highest field, yellow is negligible field).  Overlaid on the map are the RF cavities (orange) and coil (red).
  The extraction trajectory for H$_2^+$ is shown.
  Right: Illustration of the Opera3D finite
  element magnetic model showing one quarter of the cyclotron 
  with the pole, the return yoke, and the coil.  \vspace{0.2in}
\label{injcyclo} }
\end{figure}

\subsubsection{Introduction to Cyclotrons}

It is worthwhile to briefly
review cyclotron design and terminology, as we will refer to this information in the text 
that follows.     

A compact cyclotron has a monolithic magnet and 
circular coil.  
The alternative to the compact design segments the magnet,
leading to a ``separated-sector design'' such as is used in the Paul Scherrer Institut (PSI)
Injector II (3~mA of protons at 72~MeV).  Typically, the
compact design is most economical at lower energies ($\lesssim 100$~MeV), but is less
flexible than a separated-sector design.   The dipole magnet in either case
may be resistive, as with IsoDAR, or super-conducting.

In a compact cyclotron, particles  
from the ion source are injected axially at the
center, in our case via a 
``spiral inflector'' which directs the beam from a vertical direction
to the horizontal median plane.   
An RF cavity system accelerates the
particles and, as the beam gains energy, the trajectory is bent
 by a dipole magnetic field.  As a result, particles follow spiral orbits with 
radius increasing with energy. The spatial separation between the ``turns'' grows smaller
as the beam approaches the outer edge of the cyclotron.

Achieving more than 10 to 20~MeV/amu requires an ``isochronous design"
to keep the beam revolution frequency synchronized with the RF field. 
In an isochronous design, the time for one revolution
in the cyclotron is independent of particle energy.  

For vertical focusing and to facilitate beam stability, the pole faces
are formed into  pie-shaped wedges alternating a narrow pole gap (``hill") 
region with a larger gap (``valley") region. In Fig.~\ref{injcyclo}, the
high field regions in magenta correspond to the hills.    RF cavities are typically placed
in the valley region (yellow in Fig.~\ref{injcyclo}) that has
negligible field.   The RF cavities are overlaid in
Fig.~\ref{injcyclo} in orange. The IsoDAR cyclotron has 4-fold symmetry 
in this magnet pole configuration.

\subsubsection{Specifics of the Design}

The required experimental parameters (see Table~\ref{assumptions}) are
achievable with the IsoDAR base design, shown in Fig.~\ref{injcyclo} and described
in Table~\ref{tab:injtab}.  This design has been developed as the DAE$\delta$ALUS Injector Cyclotron (DIC), with technical details provided in Ref.~\cite{NIM}.

Fig.~\ref{injcyclo}, left, provides the magnetic field map of the
present design, where  the hill has 2~T magnetic field and the return
yoke (cyan) has -1.5~T.  A 3-D rendering of one sector is shown on the right side of the figure. The single circular coil associated with a compact cyclotron is indicated in red on both figures.

Throughout acceleration,
the isochronism accuracy, the degree to which the particle revolution
frequency matches the RF frequency, in this initial physics model is better than $5.0
\times 10^{-4}$ and the phase diagram is maintained in a narrow  range
($\sim \pm 4^\circ$).  Beam quality is not diminished by 
resonance crossing~\cite{resonance}, since this occurs quickly.   A second resonance crossing is observed near the extraction region, but accurate beam dynamics simulations have shown that its effect is negligible. 

Other features of the design, including RF frequency, harmonic number, and the average field value (see Table~\ref{tab:injtab}), are selected
to match the DAE$\delta$ALUS SRC.    This potential future use of the
IsoDAR cyclotron adds to the cost-effectiveness and does not diminish
the design in any way.
The large hill gap of 10~cm allows ample space for the beam envelope,
and provides good conductance for the $< 10^{-7}$~mbar vacuum required to minimize beam loss due to interactions with residual gas.  Vacuum pumping will be provided by eight cryopanels located in the valley regions, possibly integrated with the RF cavities.
The angular width of the hill, in the range $28^\circ$ to $40^\circ$, 
allows for adjustments to optimize isochronism and vertical focusing.  
All of these parameters will be studied to provide an optimal design. 

\begin{table}[t]
\centering
{
\begin{tabular}{lrrllrrl}
\hline
$E_{max}$ & 	60~MeV/amu	&	$E_{inj}$ & 35 keV/amu \\
$R_{ext}$ &	1.99 m		&		$R_{inj}$ &55 mm  \\
$<B>$ @ $R_{ext}$ &1.16 T	 &	$<B>$ @ $R_{inj}$ &	0.97 T  \\
Sectors		& 4			&		Hill width	&	28 - 40 deg \\
Valley gap	& 1800 mm	& Pole gap	& 100 mm  \\
Outer Diameter & 6.2 m	 & Full height & 2.7~m  \\
Cavities	& 4					& Cavity type	& $\lambda/2$, double gap  \\
Harmonic &	 6th		&			RF frequency	& 49.2 MHz  \\
Acc. Voltage	& 70 - 240 kV	 & Power/cavity &	$310$~kW  \\
$\Delta E$/turn	 &1.3~MeV	& Turns &95  \\
$\Delta R$ /turn @ $R_{ext}$	& $>14$ mm	 & $\Delta R$/turn @ $R_{inj}$ & $>56$ mm  \\
Coil size & 200x250 mm$^2$ & Current density	 & 3.1 A/mm$^2$  \\
Iron weight & 450 tons	& Vacuum  & $< 10^{-7}$ mbar  \\
\hline
\end{tabular}}
\caption{  Parameters of the DAE$\delta$ALUS injector
  cyclotron, from Ref.~\cite{NIM}.}
\label{tab:injtab}
\end{table}

\subsubsection{Results of Simulation}

Within the DAE$\delta$ALUS design effort, extensive and precise simulations targeting the most challenging aspects of
high power hadron drivers have been pursued.  These studies, mainly of
stationary distributions and losses, 
are reported in Refs.~\cite{NIM, SpaceCharge}.

The beam dynamics model is based on the OPAL (Object Oriented Parallel Accelerator Library) software framework~\cite{OPAL}. 
The model is validated using measured data from the PSI high power
Ring Cyclotron ($1.4~\mathrm{MW} \approx 590~\mathrm{MeV} \times 2.4~\mathrm{mA}$ continuous wave)~\cite{Bi}.

The main conclusions from the study relevant for IsoDAR (from Ref.~\cite{SpaceCharge}) are:
\begin{enumerate}
\item The simulation shows that beam transport in the cyclotron is
  space-charge-dominated. Interestingly enough, however, these space-charge forces are high enough to yield longitudinal stability of the bunch.
\item No flattop cavity is (therefore) required and the four valleys are available for installing the accelerating cavities.
\item For an extraction energy of 60~MeV/amu, the cyclotron needs 106 turns (comparable with the 72~MeV Injector II at PSI).
\item The  H$_2^+$ beam can be extracted with beam loss on septum of less than 150 W for 100\% duty cycle,  which will not result in significant issues at extraction (the last turn separation is 20~mm; for comparison the PSI Injector II has 20.5~mm).
\end{enumerate}

\subsubsection{Extraction}

\begin{figure}[t]\begin{center}
\includegraphics[width=5.5in]{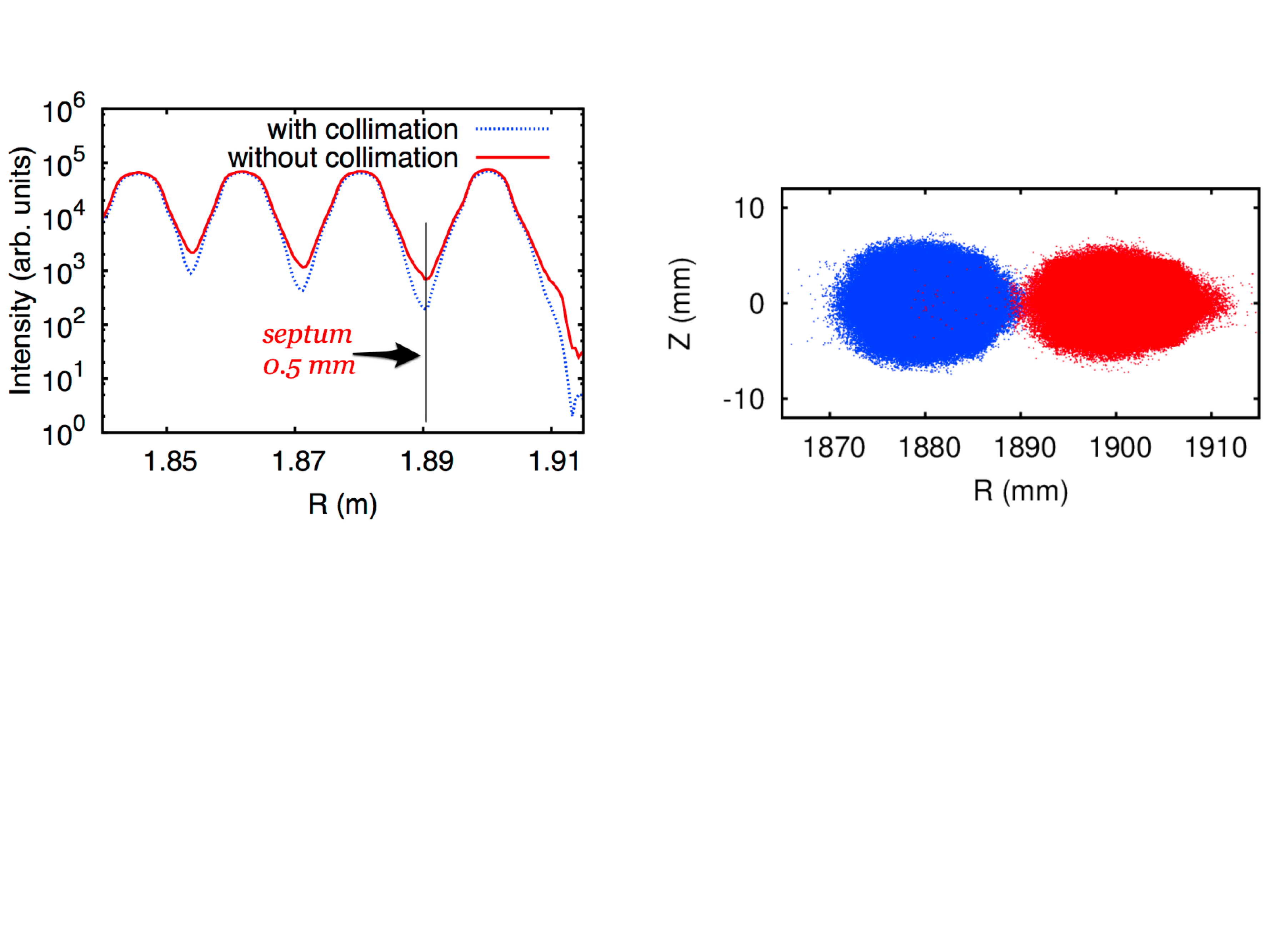}
\end{center}
\vspace{-2.1in}
\caption{Object Oriented Parallel Accelerator Library simulations for 5~mA of H$_2^+$ beam in the DAE$\delta$ALUS Injector Cyclotron.  With proper beam forming collimation in the central region, less than 1\% of the beam will be intercepted by a 0.5~mm septum demarking the extraction channel.  The figure on the right shows the next-to-last (blue) and last turn (red) as it enters the extraction channel. The overlap of the final turn with the preceding one is small as the beam halo is minimal. The figure is from Ref.~\cite{SpaceCharge}.
\label{dic_turn_sep} }
\end{figure}

IsoDAR's baseline plan for extraction is to use an electrostatic septum. As described in the previous section, the last turn separation of 20~mm allows for such a system to extract the beam with low losses. Figure~\ref{dic_turn_sep} shows the results of beam simulations, demonstrating that less than 1\% of the beam is lost on a 0.5~mm septum.

Furthermore, as H$_2^+$ ions are employed, a narrow and thin stripper foil can be placed upstream of the septum which will intercept all particles that would hit the septum. The resulting two protons from the ion will spiral inwards and avoid the septum altogether.  These can be collected with a suitably-placed catcher which may be made to take substantially more heat than the septum.

Thus, the electrostatic-septum extraction system is seen as low risk and allows for synergy with future physics projects, such as injection for the larger DAE$\delta$ALUS SRC and with the medical industry, as we discuss below. This can therefore be considered a practical and cost-effective solution.

Foil-stripper extraction may seem advantageous over electrostatic extraction because of the preponderance of stripping in modern commercial cyclotrons in this energy range. These machines, such as the Cyclone 30 IBA~\cite{cyclone} and TR-30 EBCO~\cite{TR30EBCO}, accelerate 30~MeV H$^-$ beams which, after stripping, are bent in the opposite direction to easily leave the cyclotron.  The two protons (after stripping the H$_2^+$) will spiral inwards and, because of the hill/valley field variations, an extraction channel could be designed to bring the beam out. However, care is needed in order to avoid the central region with the axial injection components. Foil lifetime in IsoDAR's very high currents would be a serious consideration.  The mean life of the IBA and EBCO stripper foils is 20-40~mA$\cdot$h. That is, the foil lasts 20-40~h~\cite{PrivKuo} with a beam current of 1~mA. The 70~MeV Legnaro cyclotron proposes to use a 6-stripper carousel to achieve a continuous operation of 14~days, with a maximum beam current of 0.7~mA.  However, foil lifetime is substantially shortened at higher beam currents, and with a single stripper station it is unlikely that a foil could be found that would have a practical lifetime when exposed to a 5~mA beam.

\subsubsection{Assembly  Underground}

Transport and assembly of an accelerator underground presents 
engineering challenges.  In the case of IsoDAR, the cyclotron weighs around
500~tons and is approximately 5~m in diameter when fully assembled.     Larger or more massive designs are likely to be unfeasible.
Access drifts to the underground space presents an important
constraint.   Using KamLAND tunnels as an example,  the narrowest 
size is 2.7~m in width and 3.2~m in height~\cite{ShimizuPriv}.   
The minimum diagonal length is therefore 4.2~m.  This constraint
requires a design that can be broken down into smaller pieces.    

The IsoDAR base design meets this requirement.
The iron pieces are readily accommodated.  The yoke and return slabs
do not have dimensional criticality, so can be bolted or welded once
transported to the cavern.  Pole pieces are a little more challenging as
each of the hills must be machined as a discrete part in order to minimize
field inhomogeneities from dimensional variations.   The welds must be
made in the valleys where the field is negligible.   In our design, each hill is a
wedge of radius approximately 2~m, arc length of approximately
1.5~m, and thickness less than 1~m. These should present no
transport problems.  The pole base, to which the hill sections are
bolted, will have an assembled diameter of 4~m.  To reduce total
weight, the base can be built up from several disks of 15-30~cm
thickness, which will be brought into the hall tilted diagonally. 
Most other components of the cyclotron are not an issue. 
Each of the RF cavities will fit into the valley sections
of the poles and transport is therefore is not a problem.  
The vacuum chamber must encompass the entire inner region of the cyclotron, 
but can be made from sections welded together or using pole
pieces themselves for vacuum surfaces. 

The primary underground assembly issue is the coils.
The inside radius of each of the two coil packages is 4.1~m
and the cross section
of each coil package is 20$\times$25~cm, so that the outer diameter is 4.5~m.  This means that an intact coil cannot be brought underground,
even when tilted on the diagonal.    Two alternative solutions for
assembling the coils are now under study.    First, it is possible to
wind in place, but this would require transport and assembly of substantial
tooling equipment, and so appears to be least cost-effective.    The
favored alternative is segmenting the coil.    To understand this technique, we an look to the 
TRIUMF cyclotron, which has a coil constructed from six segments.
The TRIUMF conductor is made from
aluminum plates approximately 2~cm thick and 60~cm high, with 18 such
plates bundled together to make a segment of the coil.  At the ends of
each of the six segments, each plate is welded to its corresponding
plate in the adjacent section.   The result is a single monolithic
coil which, in the case of TRIUMF, is 18~m in diameter.   Our case is 
significantly less challenging.  
Dividing the coil into two segments would be sufficient for transport
 into the cavern, so only two weld sections would be needed. Power supplies that provide 15~kA allow only 10~turns for the
necessary 150~kA.  Welding 10~turns is feasible, low risk,  
and would be cost effective.

Care must be taken in planning the staging, assembly and
rigging of the device underground.  The heaviest pieces are expected to be less than 50~tons.  
Several options can be explored to minimize assembly issues, and
possibly ease the requirements for cranes.  One would be the
building of a jacking structure capable of lifting the entire upper
pole and yoke assembly, and then possibly sliding it in place over the
lower yoke/pole.  This structure, similar to one installed at TRIUMF,
would also provide a way of efficiently splitting the magnet steel for
access to the mid-plane.  This access is extremely important, and must
be provided.
Another option would be to mount the entire cyclotron in the vertical
plane.  
The pole pieces could then be mounted on substantial rails, and
splitting accomplished by sliding the magnet halves apart along the
rails.  The geometry of the return yoke is fairly arbitrary, and does not
need to be circular.  A square configuration could provide
adequate steel for containing the flux, and would facilitate mounting
and support of the cyclotron in a vertical 
orientation. A vertical orientation for the cyclotron plane also eases another
problem:  the axial injection line will require about 4~m of
space between the ion source and the spiral inflector at the cyclotron
mid-plane.  If the cyclotron plane is horizontal, this axial injection
line must be vertical, or must have a 90$^\circ$ bend, complicating the
optics and transport of the intense beams from the ion source.
With a vertical cyclotron plane, the axial injection line is
horizontal, and can be accommodated with much greater ease.

In all, while presenting some interesting engineering challenges, 
the transport and installation of the IsoDAR cyclotron underground is feasible.

\subsubsection{Duty Factor, Beam Structure Requirements and Total Availability}

The half-life of $^8$Li is 841~ms. Therefore, the experiment does not benefit from having a machine with a narrow pulse time structure.
With regard to time structure, the IsoDAR
cyclotron can be considered a CW source and is therefore a good match to the requirements.

We assume 90\% beam availability to reach the physics goals.  This is set
by the need for beam-off data in order to measure backgrounds.  Since there
are no beam-structure requirements, the 10\% downtime can be scheduled as needed to
accommodate maintenance. PSI recently achieved a
week with 99.95\% up-time~\cite{PSIprivate} and the TR30 cyclotron 
(30~MeV, 1.2~mA) averages 98.3\% uptime \cite{TR30up}.

\begin{figure}[t]\begin{center}
\includegraphics[width=3.5in]{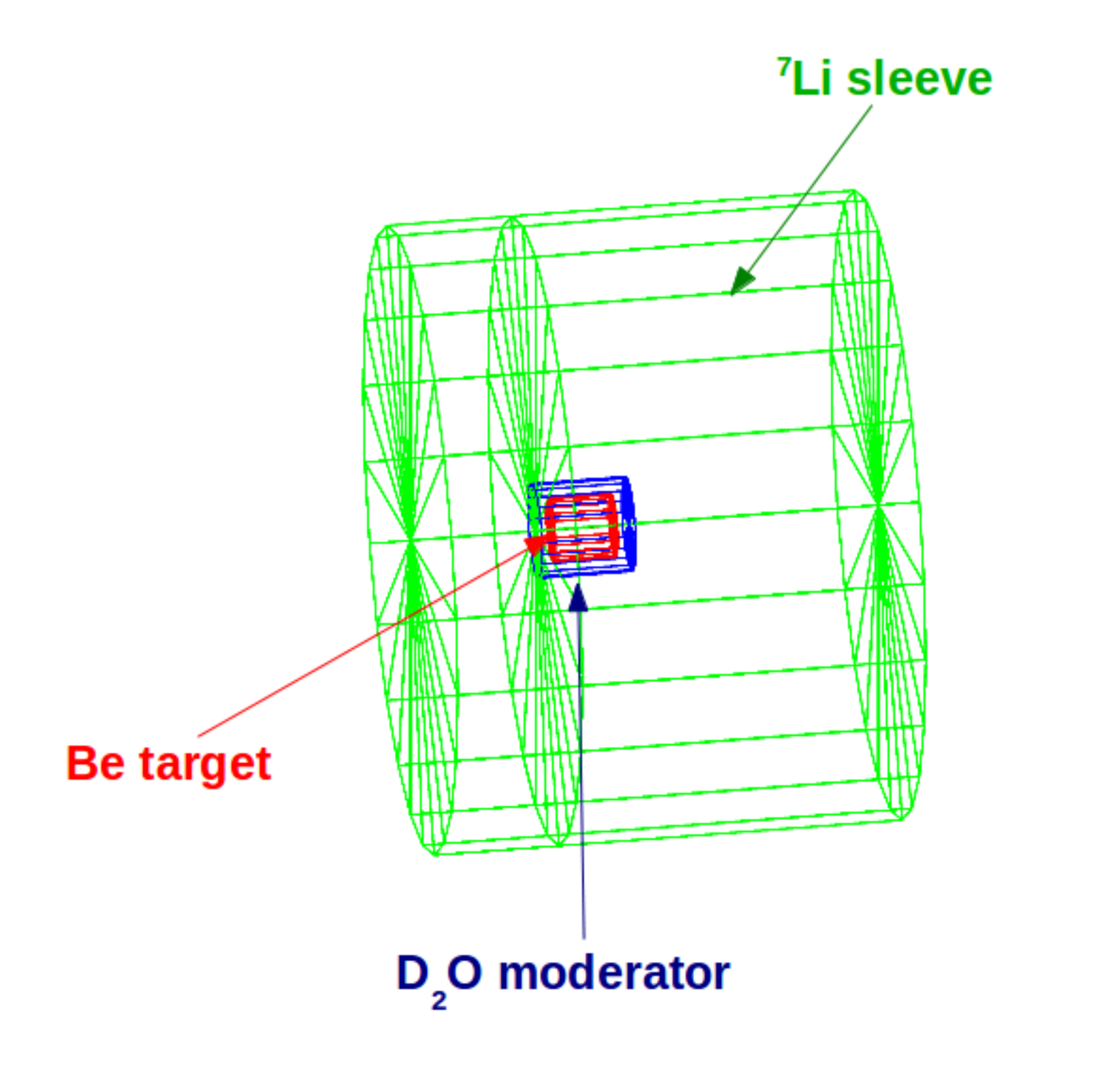}
\end{center}
\vspace{-.4in}
\caption{The beryllium target surrounded by the $^7$Li sleeve. The sleeve is 150~cm long and has a 200~cm outer diameter.
\label{geometry} }
\end{figure}

\subsection{The Beam Transport Line\label{transport}}

The 10~mA of 60~MeV/amu beam will be transported from the cyclotron to a 
target system designed to maximize the $\bar \nu_e$ flux from isotopes
that decay-at-rest with high $Q$-value.  
In consideration of target cooling and degradation and given the 600~kW beam
power,  we require a uniform beam distributed across the 20~cm
diameter target.  This can be easily accomplished with wobbler
magnets:  two orthogonal dipoles excited sinusoidally (and 90$^\circ$
out of phase) 
that sweep the beam in a circular pattern.  
Adequate uniformity and sharp falloff can be obtained using 
several concentric circular sweeps, from varying the sinusoidal
driving voltage.  Such a system can achieve 
$\pm 2\%$ uniformity over a 30~cm diameter field~\cite{wobble},
which is substantially larger than is required for IsoDAR.   Spreading the beam in this fashion reduces 
the power density on the target to about 2~kW/cm$^2$.      As this is
a CW source rather than a pulsed beam, there are no risks of shock
heating or thermal ratcheting.   

\subsection{The Antineutrino Source}
The $\bar \nu_e$ source consists of a beryllium target surrounded by a
99.99\% isotopically pure $^7$Li sleeve. 
There is a 5~cm layer of heavy water moderator 
in between the target and sleeve. This geometry is shown in Fig.~\ref{geometry}. We provide details of the base target and sleeve designs in the following section.

\subsubsection{The Target \label{target}}

The main purpose of the target is to produce a large number of neutrons through the
interaction of the beam with beryllium.  Beryllium has a high neutron production rate because the neutron binding energy is only 1.66~MeV.
Although the protons 
range out in $<3$~cm, a target length optimization study  
finds that secondary neutron interactions
contribute significantly to the total $^8$Li (and resulting $\bar \nu_e$) rate. 
We have optimized the length of the beryllium target to 20~cm, leading to a 10\% $\bar \nu_e$ flux contribution from the target alone. 

The energy deposition ($dE/dx$) in the target varies as roughly $1/E$, 
with the Bragg peak at $\sim 2$~cm.   This places the highest heat
deposition at the end of the proton range, where proton energy is low and neutron production is small. Pure beryllium is therefore not required here.    
In response to this,  we are considering a segmented target, 
with a disk of BeO at the region of the Bragg peak.  A ceramic that is
easily available commercially, BeO has a high thermal conductivity  (330~W/m$\cdot$K) and melting point
2507$^\circ$C. We note, however, that BeO cannot be used for the entire
target because the oxygen substantially diminishes neutron production.

An alternative is to design for a beryllium target that
can handle the IsoDAR power deposition of 2~kW/cm$^2$.   
A beryllium target for boron neutron capture therapy was demonstrated~\cite{MITThesis} at 
MIT-LABA that could withstand 6~kW/cm$^2$, using
a water-based cooling system at the back of the disk.
While this design has been deemed feasible, we consider it riskier than the
design with the BeO insert at the Bragg peak region.    

\begin{figure}[tb]
  \begin{center}
    \subfigure[Energy spectra of neutrons exiting the target and entering the heavy water moderator (red), exiting the moderator and entering the sleeve (blue), and of neutrons reflected back into the moderator from the sleeve (green).  ]
                     {\label{fig:l2ea4-f01}\includegraphics[width=80mm, height=57mm]{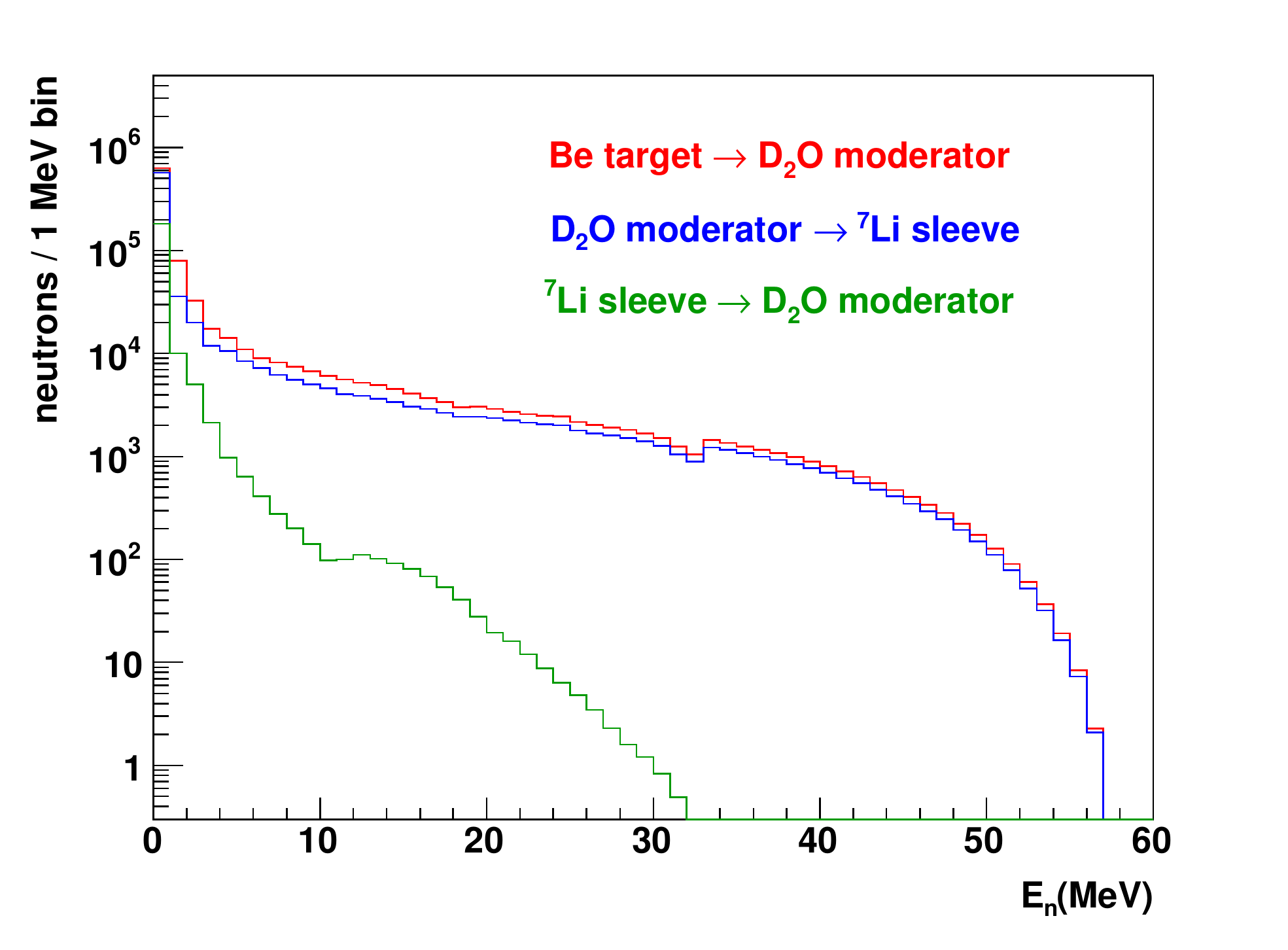}}
\hspace{1mm}
    \subfigure[Energy spectra of neutrons exiting the sleeve and entering the reflector (red) and neutrons reflected back into the sleeve (blue).]
                     {\label{fig:l2ea4-f02}\includegraphics[width=80mm, height=57mm]{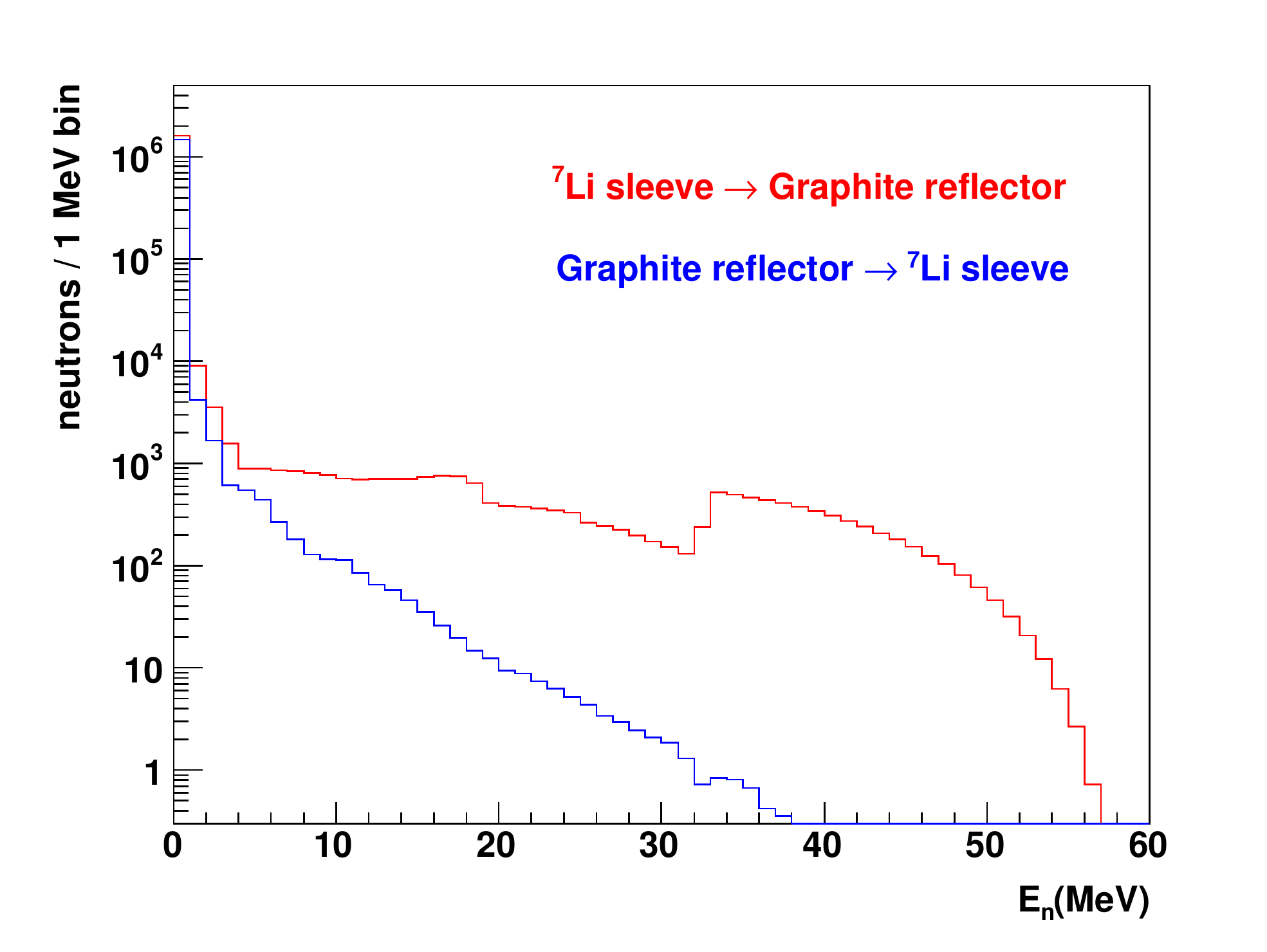}}\\
  \end{center}
  \vspace{-.2in}
  \caption{Energy spectra of neutrons as they pass the most relevant geometric boundaries from one volume to another.}
  \label{fig:l2ea4-f0}
\end{figure}

D$_2$O is used for target cooling because interactions of fast neutrons from the target inside of this volume increase the total neutron flux.   This also
provides moderation of the fast neutrons before they enter the sleeve.

Beryllium and BeO are light targets and therefore produce mostly
short-lived isotopes upon activation.     
Using low-A materials reduces the risk related to
removal and storage of the activated target.   We discuss this further 
when considering alternative target materials below.
The radioactive target handling design will be based on the successful MiniBooNE
horn and target handling system.    As already stated, the volume 
that receives the primary physical stress and dose is small and the coffin system can therefore be expected to be much more compact than MiniBooNE's.

The target and sleeve must be engineered for the highest possible reliability.  The goal is for the target to last for the full length of the experiment and not have to be changed during the run.  Infrastructure for handling the highly radioactive target would be very difficult to integrate into the confined area allocated for it, and changing a target would inevitably require a substantial down-time for the experiment.

\subsubsection{The Sleeve \label{sleeve}}

A sleeve surrounding the neutron source target produces the majority of the
$\bar \nu_e$ flux. The 150~cm long, 200~cm outer diameter cylindrical
sleeve surrounds the target and D$_2$O layer. This volume is embedded
40~cm into the upstream end of the sleeve; a window allows the beam
to impinge on the target.  The sleeve, composed of 99.99\% isotopically
pure $^7$Li, utilizes thermal neutron capture and
the subsequent beta-decay of the resulting $^8$Li for $\bar \nu_e$
production. A neutron reflector, composed of graphite and steel,
surrounds this volume. The geometry of the sleeve and the position of
the target inside have been chosen to maximize $\bar \nu_e$ production
in consideration of, among other things, the KamLAND tunnel space restrictions. 

\begin{figure}[t]\begin{center}
\includegraphics[width=3.8in]{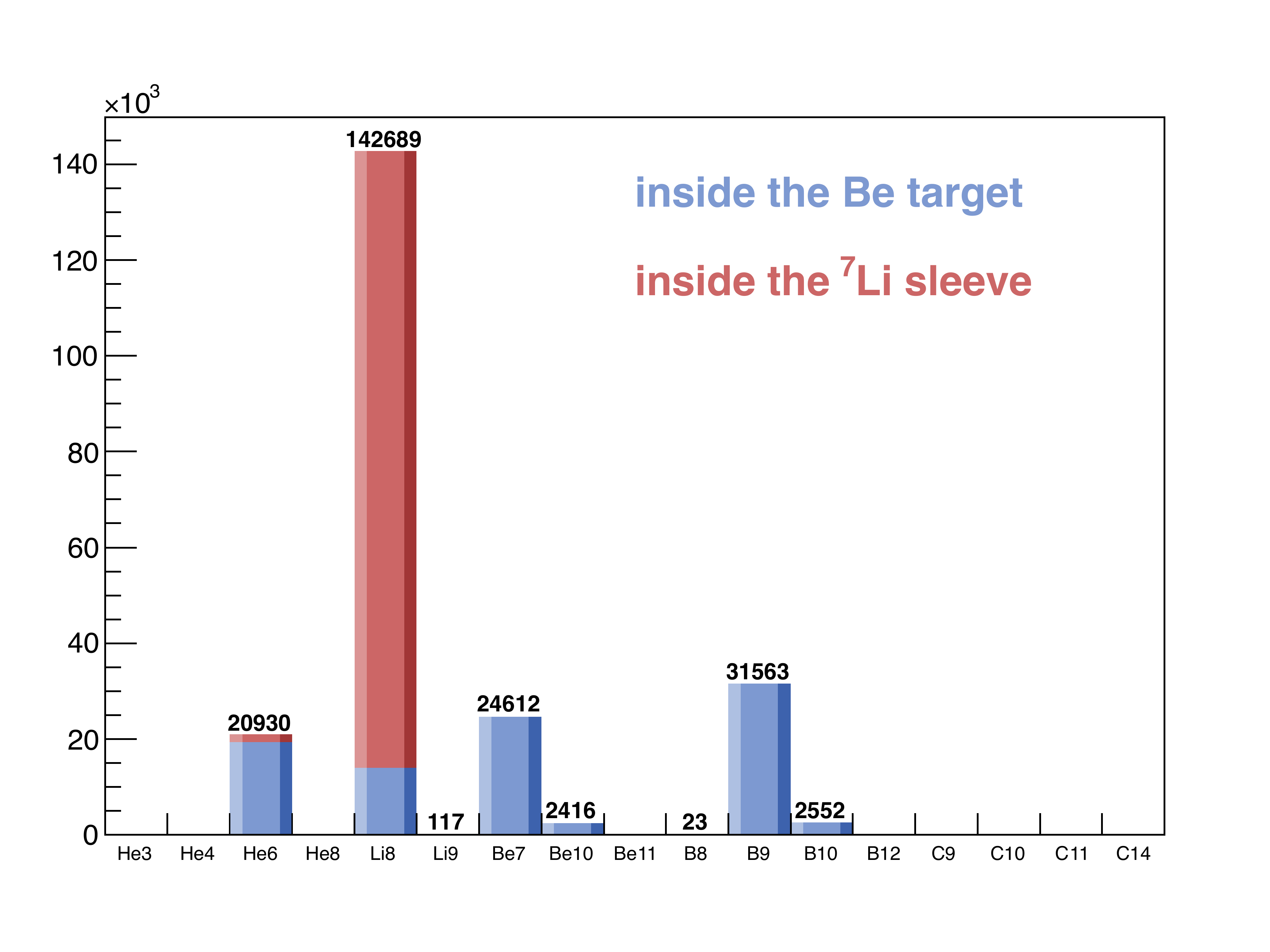}
\end{center}
 \vspace{-.3in}
\caption{Isotopes produced inside the target and sleeve geometry for 10$^7$ protons on target, given the baseline IsoDAR design and 99.99\% $^7$Li target sleeve purity.
\label{isotopes} }
\end{figure}

The energy spectra of neutrons exiting the target as well as exiting the sleeve
can be seen in Fig.~\ref{fig:l2ea4-f0}. Reflected neutrons at both geometric thresholds can
also be seen in the figure.   Neutrons which thermalize and subsequently capture
produce $^8$Li. Fig.~\ref{isotopes} shows the isotopes produced 
for $10^7$ protons on target at 60~MeV. The isotope production rates shown set the benchmark for comparisons of alternative target and sleeve materials below.

Isotope production rates are based on software simulation studies with GEANT4~\cite{G4}.  GEANT4 provides a large set of hadronic
models (data-based, parameterized, and theory-driven), each model being
defined for a given type of interaction within a specific energy range. For this particular application, the ``QGSP-BIC-HP" physics
package was chosen. The most relevant physics model for our study is the
pre-compound nuclear implementation utilized by the Binary Cascade (BIC)
model. The following hadronic processes are included: inelastic
scattering, elastic scattering, neutron fission, neutron capture, lepton-nuclear
interactions, capture-at-rest, and charge exchange. Neutrons with energy below 20~MeV are treated with the high-precision (HP) model and the 
ENDF/B-VII~\cite{CSEWG} data library.   The low-energy GEANT4 software has been
well-benchmarked due to use in medical physics~\cite{bench}.

\subsubsection{Shielding}

The target, sleeve, and graphite reflector are embedded in a steel and concrete shield.
We assume 3.5~m of shielding here.   Table~\ref{attenlen} provides the attenuation
lengths of concrete and iron at neutron energies of $<$25~MeV and $>$25~MeV. This volume 
meets the required rates for occupancy and also provides 
sufficient shielding to address issues of activation in the mine. The shielding is also adequate for eliminating neutrons that can mimic an antineutrino signal in the detector itself.  Those isotopes that are produced in the shielding represent a negligible contribution to the IsoDAR source above $\sim$3~MeV.
As can be seen in Fig.~\ref{fig:l2ea4-f0}, the majority of neutrons that exit the sleeve are ``fast" as the volume is well-designed to absorb thermal neutrons.

\begin{table}[t]
\centering
{
\begin{tabular}{|lc|cc|cc|}
\hline
\multicolumn{2}{|c|}{~~}&\multicolumn{2}{|c|}{$<$25
 ~MeV}&\multicolumn{2}{|c|}{$>$25~MeV}\\ 
Material & density (g/cm$^3$)& g/cm$^2$  &~cm & g/cm$^2$  &~cm\\\hline
Iron & 7.8 &100 & 12.8 &  138 & 17.7\\
Concrete & 2.4 & 40 & 18.8 & 65 & 27.1 \\ 
\hline
\end{tabular}}
\caption{Attenuation lengths for neutrons in iron and concrete.}
\label{attenlen}
\end{table}

The compact shielding volume is important within the context of the experiment.
If IsoDAR is run at KamLAND, then the source would be placed in a 
space that presently houses a control room. This room would be relocated, allowing the necessary proximity, including shielding, to the KamLAND detector.

\subsection{Flux and Event Distributions}

\begin{figure}[t]
\begin{center}
\includegraphics[width=4.in]{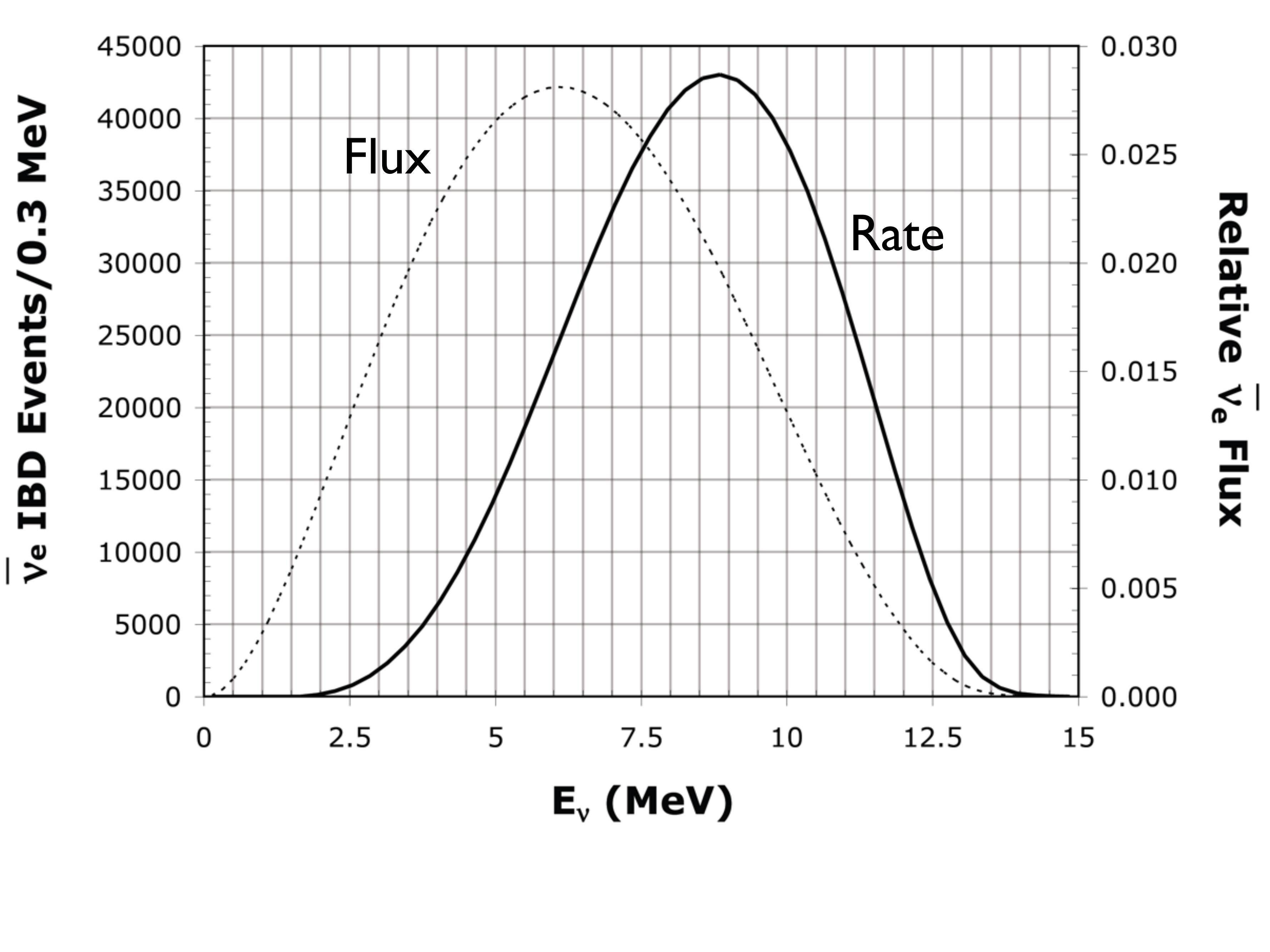}
\end{center}
\vspace{-.6in}
  \caption{The flux and IBD event distribution expected with the baseline IsoDAR design.
  \label{flux}}
\end{figure}

Fig.~\ref{flux} shows the flux and event rates from the $\bar
\nu_e$ source expected in KamLAND (897~ton fiducial volume), with detector
center located 16~m from the target face. In 5~years, nearly one million IBD events and 7200~$\bar \nu_e$-electron scatters will be detected.
The physics allowed by this event sample  is discussed in
Ref.~\cite{PRL}. Any alternative design must meet these rates in a cost-effective manner.

\section{Partnerships Allowed By The IsoDAR Base Design \label{widerimplications}}

The design of the IsoDAR accelerator has been influenced by potential  
partnerships with industry and with other particle physics
experiments.
This is essential to the cost-effectiveness of the experiment.  Thus,
when weighing the design choices, we have kept in mind the impact
of design changes on partnerships. Here, we consider the potential needs of these partnerships.

\subsection{With the Medical Isotope  Industry \label{industry}}

IsoDAR opens the opportunity for a partnership between neutrino
physics and industry, an aspect that adds considerably to the cost-effectiveness
of the base design. Cyclotrons are widely used to produce medical isotopes.
A 60 to 70~MeV machine produces a unique set of isotopes not available
at lower energies, summarized in 
Table~\ref{tab:med}.
The latest generation of accelerators at 70~MeV, running at 750 $\mu$A,  are sold by 
IBA~\cite{IBA} and BEST~\cite{BEST}.  IsoDAR's 10~mA of protons will lead to a substantial increase in production of these isotopes.  Ref.~\cite{JoseIsotope} provides a tutorial on isotope production and its connection to IsoDAR. 

\begin{table}[th]
\centering
{
\begin{tabular}{|l|c|c|}
\hline
Isotope & Half-life & Use \\ \hline
$^{52}$Fe & 8.3 h &  The parent of the PET isotope $^{52}$Mn  \\
   & & and iron tracer
  for red-blood-cell formation and brain uptake studies.\\  \hline
$^{122}$Xe & 20.1 h &  The parent of  PET isotope $^{122}$I used to study
brain blood-flow. \\ \hline
$^{28}$Mg  & 21 h & A tracer that can be used for bone studies,
analogous to calcium. \\ \hline
$^{128}$Ba  & 2.43 d & The parent of positron emitter $^{128}$Cs. \\
  & & As a
  potassium analog, this is used for heart and blood-flow imaging. \\ \hline
$^{97}$Ru & 2.79 d & A $\gamma$-emitter used for spinal fluid and liver
studies. \\ \hline
$^{117m}$Sn & 13.6 d & A $\gamma$-emitter potentially useful for bone
studies. \\ \hline
$^{82}$Sr & 25.4 d &  The parent of positron emitter $^{82}$Rb, a
  potassium analogue.   \\ 
& &  This isotope is also directly used as a PET
  isotope for heart imaging. \\ 
\hline
\end{tabular}}
\caption{Medical isotopes relevant at IsoDAR
  energies, from Ref.~\cite{cost-cycl-2005}. }
\label{tab:med}
\end{table}

Several national research laboratories -- e.g. ORNL and BNL in the US, INFN-Legnaro in Italy, and ARRONAX in Nantes, France -- are seeking to acquire, or are actually using high-power cyclotrons in IsoDAR's energy range for isotope production.  ARRONAX has installed the first IBA C70 (0.75~mA, 70~MeV proton cyclotron)~\cite{C70} and ORNL is requesting funds for a similar machine~\cite{Holifield}.  Furthermore, BNL conducted a DOE-funded study with Jupiter~\cite{cost-cycl-2005} for a machine in this energy range and Legnaro is acquiring a 70~MeV, 1~mA cyclotron from Best Cyclotron Systems~\cite{Legnaro}.

The programs to be pursued at all of these laboratories involve research with isotopes, ultimately for medical or industrial applications.  Legnaro intends, in addition, to collect and accelerate selected ions in a RIB (Radioactive Ion Beam) facility.  Target development is a primary goal in their research programs. The delicate process of extracting desired isotopes places unique demands on targets, with a limiting factor always being the power that can be absorbed.  Research to increase the power-handling capabilities of targets--enabled by the higher beam currents from IsoDAR-class machines-- will ultimately lead to increased yield and improved efficiency in isotope production.

In addition, the high currents can be used for exploring methods for sharing beam between many target stations.  This would provide for an overall increase in product output, or versatility in simultaneous production of various isotopes at different target stations.

An elegant solution for such beam sharing is proposed for an H$_2^+$ beam extracted with a conventional electrostatic septum.  The transport line from the extraction point includes a set of focusing elements that form the beam into a long horizontal shape.  A stripping foil intercepts a small amount of the beam at one of the lateral edges, the amount intercepted depending on how far in the stripper is moved.  Downstream of the foil will be protons for the ions going through the foil, and H$_2^+$ ions that have missed the foil.  This beam is passed through a dipole magnet that bends the protons more strongly into a separate channel where they are focused and transported to a target.  

The remaining H$_2^+$ ions are passed through another focusing element that again produces an elongated beam at the site of a second stripper, with a dipole behind it to again separate protons to be directed to the second target. This process can be repeated as often as desired, for 1~mA per target in principle as many as ten times.  

\subsection{In Particle Physics}

As has been shown for the IsoDAR experiment, 60~MeV proton beams can be prolific sources of neutrinos through the production of beta-decaying isotopes.  While IsoDAR is the first of this class of experiments, undoubtedly others will be proposed in the future.  Accelerators for such new experiments can build on the base of experience gained with IsoDAR to provide performance tailored to their needs.

Perhaps more relevant, the IsoDAR cyclotron can also be used in a chained-cyclotron system for applications in particle physics. IsoDAR was originally conceived as an injector for the DAE$\delta$ALUS system, which uses ``accelerator modules" with sub-units consisting of an ion source, an injector cyclotron that accelerates to 60~MeV/amu, a SRC that accelerates to 800~MeV, and a target/dump for pion production~\cite{EOI}. In a chained cyclotron system, there is value in accelerating H$_2^+$ in the injector, and extracting the ion electrostatically. This means that the intact H$_2^+$ can be extracted at high energy using multiple foils and provides a very clean extraction mechanism at 800~MeV. In principle, H$^-$ can also be extracted by stripping foils, but is not considered a candidate because Lorentz dissociation sets in well before 800~MeV.

\subsection{ADS Technology}

High power accelerators are of interest to the nuclear reactor
community for the purpose of accelerator driven systems (ADS) for
thorium reactors and actinide incineration.    The interest in thorium 
reactors has risen worldwide since the Fukushima accident in Japan.
These reactors are inherently safe because they produce fission
without achieving criticality, so that when the driver shuts down, the 
reactor turns off.   ADS requires very high power and very high reliability.    Thus, an attractive system which could be
cost effective is to use multiple few-MW cyclotron systems as
drivers. Cyclotrons are inherently highly reliable and, with multiple
systems, one can be brought down for maintenance while the others
continue to drive the reactor.     Cyclotrons are sufficiently
low-cost that several can be employed per reactor.
In fact, the DAE$\delta$ALUS design originated from a
cyclotron for ADS development~\cite{Lucianofirstpaper}. Note that there continues to be substantial interest in a chained cyclotron
system for this application~\cite{LucNYThorium}.

\section{Alternative Options for the Beam and Antineutrino Target \label{overallalt}}

Having defined the base design and explained the important partnership
opportunities it affords, we can now begin to consider alternatives.
In this section we begin by considering variations on the basic concept of 
60~MeV H$_2^+$ impinging on a beryllium target surrounded by 99.99\%
pure $^7$Li. We consider various beam particles, target
materials, and sleeve materials.


\subsection{H$_2^+$ versus other Beam Particles \label{otherions}}

The viable alternatives to H$_2^+$ are high energy deuterons, protons, and H$^-$.    The argument against deuterons lies in 
activation issues.   The main arguments against protons and H$^-$ and
are in added cost due to --as we will see-- the requirement for larger machines in order to mitigate
space-charge effects.  Opportunities for 
partnerships discussed above also disfavor protons and H$^-$ compared to 
the choice of H$_2^+$.   

First, let us consider a beam of deuterons.    
Each accelerated deuteron ion would
deliver one 60~MeV proton and one 60~MeV neutron to the target.
Delivering neutrons directly to
the target, rather than relying on secondary production, seems naively attractive.   However, the problem is in slowing the neutrons so that they
productively capture to make $^8$Li, rather than escaping and activating the surroundings.    We find that 60~MeV neutrons are very difficult to
control and use successfully.   The scattering length of neutrons in
beryllium or solid lithium at 60~MeV is more than 20~cm.  
In other words, the attenuation length is large on the scale of the target/sleeve
configuration. As a result, 
it is difficult to envision a realistic target geometry for a deuteron beam which produces rates at the
level of the standard IsoDAR design.  We have explored various deuteron beam energies and have not found an optimal design
that produces the required rates.  Beyond problems with the 
neutrino source, high energy deuterons
are infamous for producing losses in accelerators that are difficult to control.
Thus, we reject a deuteron beam because there is no good technical solution 
for producing a high antineutrino flux while maintaining low activation.

As an example of a comparable proton-based cyclotron,
consider the PSI Injector II. This device has a
diameter twice the size of IsoDAR's base design and has extracted
protons of 72~MeV at 3~mA.   Can this
technology be extended to 10~mA of protons?
Based on simulations with a compact cyclotron design, one expects a significant increase of losses between
5 and 10~mA when running protons.  We expect similar effects in a
separated sector machine,
like PSI II,  although the detailed
simulation is not complete.
The increase in losses is due to space charge effects and can be
mitigated by making the machine even larger.  However, the PSI-II design, at twice
the physical size of IsoDAR's base design, would be extremely difficult to implement underground.
Another problem with proton-based machines is that the
staggered foil system described in Sec.~\ref{industry} cannot be applied for protons, and it is
unclear how high power isotope production would operate in this case.

An H$^-$ machine also presents difficulties.  Similar space-charge
issues are applicable for H$^-$ at high intensity, since the charge-per-proton
ratio is the same as for a proton in absolute value.  Therefore, a
PSI-II size design would be the minimum required.      Several other problems
arise because the H$^-$ is an ion.  First, the vacuum must be substantially 
better than required for a proton machine, and at least comparable to that of the 
H$_2^+$ machine, because of stripping with residual gas.   Second,  the H$^-$  maximum magnetic field must
be lower than 1.7~T, to avoid electromagnetic disassociation, making the H$^-$ machine even larger than the proton PSI Injector II cyclotron, which has a hill field of 2~T. 

Extraction from an H$^-$ cyclotron can be accomplished either by the same electrostatic septum system that would be employed for a proton machine such as  the PSI Injector II, or by using stripping foils.  Septum extraction would preserve the H$^-$ ion, which could be distributed to many target stations by the staggered-foil distribution system described above.  However, the usual extraction system for H$^-$ cyclotrons used in the isotope industry employs stripping foils at the outer radius of the cyclotron.  H$^-$ ions traversing these foils lose both their electrons and are bent out of the cyclotron magnetic field towards an extraction channel.  Several (three or four, depending on the symmetry of the cyclotron magnet) strippers can be placed at the outer radius, and beam shared amongst them, to distribute the total beam power to different targets. Sharing the beam on several strippers also mitigates the problem of heating in the foil, which severely affects foil lifetimes. During operation, the foils glow white-hot, perilously close to the sublimation temperature of the carbon foil material. Industry experience is that at around 1~mA a foil will last about 40~hours, an acceptable lifetime given suitable carousels to exchange foils quickly.  Higher current through a foil will decrease lifetime in a highly non-linear fashion; a factor of two will cause almost instantaneous failure.

Stripping extraction from a 10~mA H$^-$ cyclotron suitable for the IsoDAR experiment presents an insurmountable problem.  This is particularly true since only one extraction channel can be used to bring all of the beam onto the antineutrino-producing target. Thus, an H$^-$ machine would have to be septum-extracted and, as indicated above, would be at least on the order of the PSI Injector II.

Furthermore, for the application of medical isotope production using staggered foils, there
is an issue of foil lifetime.  When stripped, electrons circle back and are intercepted by
the foil. H$^-$ has a larger electron-to-proton ratio than H$_2^+$ and therefore a reduced foil lifetime.
The sum of these problems means that H$_2^+$ is preferable.

Lastly, an H$_2^+$ machine is required if the machine is to be
employed as an injector in a chained cyclotron system  
that accelerates beam beyond about 590~MeV.  This is the energy where
H$^-$ becomes impractical due to the problem of  Lorentz stripping.  
This energy is too low for ADS in thorium reactors
\cite{LucNYThorium} or for future neutrino experiments like
DAE$\delta$ALUS.  H$_2^+$ is therefore required for partnerships like these.

\subsection{Beam Energy Versus Current \label{energy}}

The basic concept behind IsoDAR physics is to flood a volume with
neutrons for creating beta-decaying isotopes and the resulting antineutrinos.  This is done with 5~mA of H$_2^+$ (equivalent to 10~mA
of protons)  at 60~MeV/amu in the base design.  However,
accelerated particle beams with a wide range of energies on various target materials can also provide a high flux of
neutrons.  One can look at options from about 200~keV to $>$GeV, from low energy D-T generators~\cite{DT} to 3~GeV
spallation sources~\cite{Japan3GeV} -- the options available throughout
the field are enormous.  It is reasonable to ask if any of these
systems could provide a viable alternative combination of energy and current to meet IsoDAR's physics goals.

The critical parameter is the $n/p$ ratio, or the number of neutrons
produced for every incoming proton
striking the target.  This ratio is highly dependent on the beam
energy.  Jongen~\cite{Jongen} has compiled $n/p$ data for low energy
beams, from D-T generators at less than 1~MeV to about 140~MeV.   For higher
energies, the ratio is addressed in spallation source studies.
It is found that, for energies of 1~GeV
and above, the ratio becomes constant for a given beam power. For example, a
2~GeV beam will require half the protons of a 1~GeV beam to generate
the same number of neutrons.  A graph of neutron production versus
beam power is presented by Pynn~\cite{Pynn} that covers energies from
about 100~MeV to several GeV.  A detailed experimental study at 62~MeV~\cite{62MeVBe} provides a specific measurement of this ratio near IsoDAR's energy.

\begin{figure}[t]\begin{center}
\includegraphics[width=5.5in]{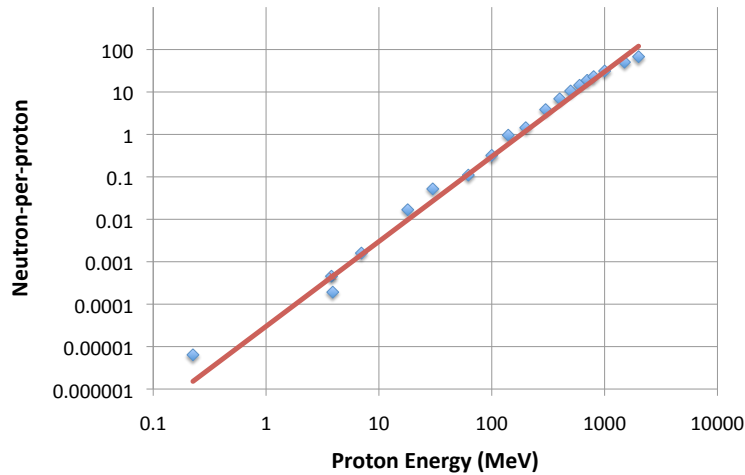}
\end{center}
\vspace{-1.2in}
\caption{Neutron yield per accelerated particle (proton or
  deuteron) versus beam energy. The data are fron Refs.~\cite{Jongen,
   Pynn, 62MeVBe}.  
\label{pnplot} }
\end{figure}

\begin{table}[p]
\centering
\begin{tabular}{lccccc}
\hline \hline													
Beam energy	&	$n/p$	&	$n/p$ (fit)	&	Req'd	&	Max.	&	Source	\\	
MeV	&		&			         &current		&	Existing	&		\\	\hline
0.225	&	6.40E-06	&	1.51875E-06	&	710	A	&	10 mA	&	D-T gen.	\\	
3.9	&	1.92E-04	&	4.56E-04	&	2.4	A	&	2 mA	&	IBA	\\	
3.8	&	4.56E-04	&	4.33E-04	&	2.5	A	&	2 mA	&	IBA	\\	
7	&	1.60E-03	&	1.47E-03	&	730	mA	&	2 mA	&	IBA	\\	
18	&	1.68E-02	&	9.72E-03	&	110	mA	&	2 mA	&	IBA	\\	
30	&	0.052	&	0.027	&	40	mA	&	0.8 mA	&	IBA	\\	
60	&		&	0.11	&	10	mA	&	10 mA	&	IsoDAR	\\	
62	&	0.11	&	0.12	&	9.4	mA	&		&		\\	
100	&	0.32	&	0.3	&	3.6	mA	&		&		\\	
140	&	0.96	&	0.588	&	1.8	mA	&	1 mA	&	IBA	\\	
200	&	1.44	&	1.2	&	900	$\mu$A	&		&		\\	
300	&	3.84	&	2.7	&	400	$\mu$A	&		&		\\	
400	&	6.912	&	4.8	&	230	$\mu$A	&		&		\\	
500	&	10.56	&	7.5	&	140	$\mu$A	&	0.2 mA	&	TRIUMF~\cite{triumf}	\\	
600	&	14.4	&	10.8	&	100	$\mu$A	&	2.2 mA	&	PSI~\cite{psi}	\\	
700	&	18.816	&	14.7	&	70	$\mu$A	&		&		\\	
800	&	23.04	&	19.2	&	60	$\mu$A	&	1 mA	&	LANSCE~\cite{lansce}	\\	
1000	&	31.2	&	30	&	40	$\mu$A	&	1 mA	&	SNS~\cite{sns}	\\	
1500	&	50.4	&	67.5	&	20	$\mu$A	&		&		\\	
2000	&	67.84	&	120	&	10	$\mu$A	&		&		\\	\hline
\end{tabular}
\caption{ Data for $n/p$ ratio versus energy, from
  Refs.~\cite{Jongen, Pynn, 62MeVBe}.  The ``$n/p$ (fit)" column presents values from a linear
  regression fit to a quadratic relationship between $n/p$ and energy.  The ``Req'd current" column lists
  the current needed at each energy to generate the flux of
 $\bar \nu_e$ required for the IsoDAR experiment.  
\label{nptable}}
\end{table}

Figure~\ref{pnplot} summarizes these data sets, with the red line being
an empirical fit to the data points.  The data sets
match well at the overlap point, including the measurement at
62~MeV,  most relevant for IsoDAR.  The
straight line (on a log-log scale) corresponds to a quadratic
relationship in $n/p$ versus beam energy over almost 4 orders of magnitude for
beam energy, with 
\begin{equation}
  (n/p) = 3\times 10^{-5} \times E^2~,
\end{equation}
 where the energy $E$ is in~MeV.

Table~\ref{nptable} displays the $n/p$ ratio as a function of energy, and assesses the beam
currents necessary for IsoDAR's antineutrino production: $2.6\times 10^{22}$ 
$\bar \nu_e$  per year with 10~mA of protons at 60~MeV.  

We have chosen 60~MeV as the ideal beam energy given the constraints of 
underground application and cost. It is clear that below 60~MeV the performance of existing accelerators falls far short of
meeting the required antineutrino flux goals.  While the higher energy ($>$500~MeV)
 machines easily meet this requirement, they will be considerably
 larger and more expensive.   This precludes
 serious consideration of higher energies for IsoDAR.

\subsection{Alternative Target and Sleeve Materials \label{EMat}}

Having determined that a 60~MeV/amu H$_2^+$ beam on target is optimal, 
we can now explore alternative target and sleeve materials. 
We begin by considering the baseline target composed of $^9$Be and compare the neutron output to heavier targets. Then, we consider materials that can be used in the sleeve for optimizing $\bar \nu_e$ production.

\subsubsection{Alternative Solid Target Materials}

The choice of target material is guided by engineering and cost requirements.   At the low-A end of the periodic table, 
it must be noted that several isotopes with $Q>3$~MeV 
can be produced at relatively high rates which decay to neutrinos, 
rather than antineutrinos, including $^8$B and $^{12}$N. 
The suppression of these isotopes is important in the choice of
materials considered here.  This eliminates carbon and some other 
potential low-A target materials. At the high-A end, we avoid
already-activated materials, such as depleted uranium, for ease of 
use.

The total $^8$Li yield calculated for various
incident proton energies and with copper and tungsten targets for comparison
with beryllium is presented below.   The alternative target materials are considered because they meet the following guidelines:
\begin{itemize}
\item High neutron yield.
\item  High melting point.
\item  High thermal conductivity. 
\item Chemically inert, low corrosion (assuming a gas coolant).
\end{itemize}

\begin{table}[h]
\centering
\begin{tabular}{lccc}
\hline \hline
beam energy/amu & $^8$Li yield (Be) & $^8$Li yield (Cu) & $^8$Li yield (W) \\ \hline  \hline
30 & 30,204 & 23,153 & 34,051 \\ \hline
40 & 49,539 & 46,028 & 75,416\\ \hline
50 & 86,333 & 75,777 & 132,151\\ \hline
60 & 144,571 & 112,004 & 206,666 \\ \hline \hline
     & Be property & Cu property & W property \\ \hline
Melting point               & 1278 $^\circ$C &   1085 $^\circ$C &
3422 $^\circ$C \\ \hline 
Thermal conductivity &  210 W/m$\cdot$K & 390 W/m$\cdot$K &  174 W/m$\cdot$K\\
\end{tabular}
\caption{A comparison of solid targets.   Top: the $^8$Li yield for 10$^{7}$ protons on target for different
  target materials and incident proton energies (considered in Sec.~\ref{energy}). Bottom: relevant material properties.
\label{target_materials}}
\end{table}

The $^8$Li (or equivalently $\bar \nu_e$) event rate is shown for the various materials in Table~\ref{target_materials}. Despite its slightly lower $^8$Li production rate, ease of handling the
radioactive target underground has led us to consider beryllium as the most attractive choice among the materials studied.
The p+W interaction will produce, in descending order of
activation, $^{179}$Ta,  $^{173}$Lu , $^{174}$Lu, $^{157}$Tb and
$^{101}$Rh~\cite{NabbiShetty}.  These are significant radiation
hazards; $^{157}$Tb  has a 100~year lifetime,
while the others are between 1 and 5~years.  Other problematic
isotopes that are among the top 20 contributors of activation from W are
$^{22}$Na, $^{55}$Fe, $^{60}$Co, and $^{85}$Kr~\cite{NabbiShetty}. 
These can be compared against the isotopes listed in Fig.~\ref{isotopes},
which are those that are produced in beryllium for $10^7$ protons on
target.  The longest lived isotope produced is $^{11}$C, with a 30~minute
lifetime, and which is produced in very small quantities.  
As a result of the added complexity of radiation handling for tungsten, 
we expect that beryllium will be the more cost effective alternative.
We note, however, that tungsten provides an alternative if power issues prove to be unsurmountable for a beryllium target.  This issue will be studied in the future.

A radical alternative is a liquid mercury target.
The primary argument for liquid mercury targets is dissipation of heat.
However, these targets, such as the mercury one produced for the SNS, are very complex and expensive, and require a large hot-cell infrastructure for routine maintenance.   The EURISOL-DS study gives cost accounting for
various MW-level targets~\cite{EURISOL}.  The associated cost of the liquid target components in the overall assembly total to
$>\$30$M, not including labor.   Also, the target itself along with the associated components requires a
large volume compared to the volume of beryllium needed for IsoDAR.  This means a
much larger radius of $^7$Li would be required. For these reasons, a liquid target is impractical for IsoDAR.

\subsubsection{Alternative Sleeve Materials}

Initially, a borated polyethelene target sleeve was considered for the
production of $\bar \nu_e$ via the neutron-capture-induced creation
and subsequent decay of beta-emitting $^{12}$B. As the neutron
inelastic cross-section on $^{10}$B is orders of magnitude higher than
the neutron capture cross-section on $^{11}$B, only $^{11}$B was included in the sleeve material. This resulted in 403,628 $^{12}$B isotopes
produced for 10$^{7}$ incident protons. However, even a small
contamination of 0.1\% $^{10}$B reduces
$^{12}$B production by a factor of $\sim$500. Recalling that stable boron is composed of 20\% $^{10}$B and 80\% $^{11}$B (and isotopic separation at the $>$99.9\% level is not trivial), this material selection was not considered further.  We also note that $^{10}$B would need to be purchased whereas isotopically pure $^7$Li is available in a substantial inventory for federal nuclear programs, and could possibly be ``borrowed" much like the depleted uranium for the D0 experiment.  

\begin{figure}[t]\begin{center}
\includegraphics[width=4in]{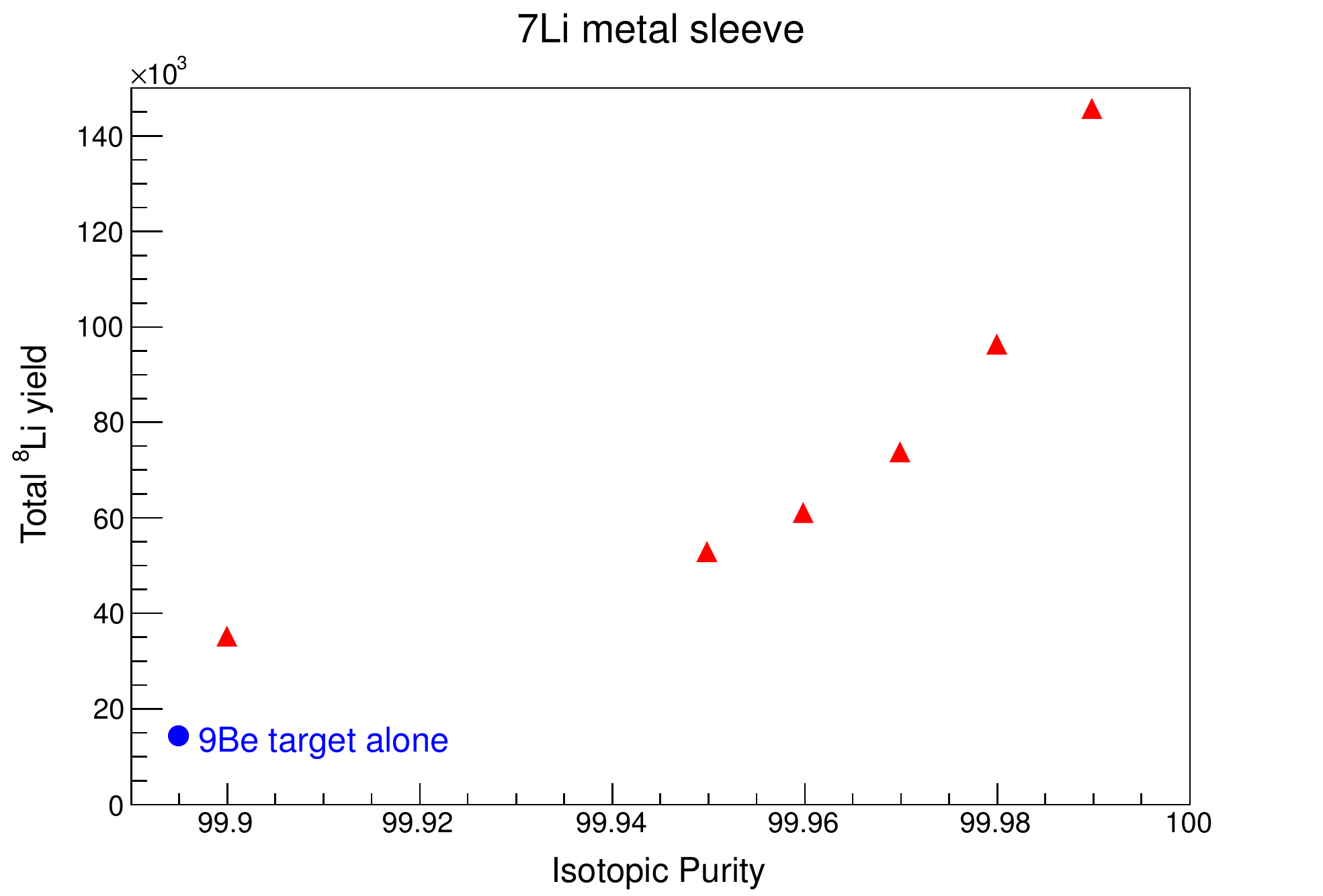}
\end{center}
\vspace{-.2in}
\caption{$^8$Li yield for various $^7$Li isotopic purities inside the target sleeve in the baseline IsoDAR design.
\label{purities} }
\end{figure}

Several variations on the choice of lithium were studied.   First, natural lithium was considered. As the neutron inelastic
cross-section for $^6$Li is much higher 
than the neutron capture cross-section on $^7$Li, only 99 $^8$Li
isotopes per 10$^7$ protons on target are actually produced inside the
sleeve with this material. 
This can be compared to the rates for 99.99\% pure $^7$Li shown in Fig.~\ref{isotopes}.  We note that the four nines in ``99.99\%" can be considered a realistic goal for the sleeve. Fig.~\ref{purities} shows the dependence of $^8$Li yield on $^7$Li isotopic purity. A LiO$_{2}$ target sleeve was also considered, but again the rates were not competitive with pure solid lithium. 

\section{Alternative Driver Designs \label{driveralt}}

In this section, we investigate alternative options of
accelerators that can be ``drivers'' in the IsoDAR system.
The machines that could possibly produce sufficient current at 
high enough energy are cyclotrons, LINACS,
FFAGs (Fixed Field Alternating Gradient), and RPSs (Rapid Cycling Synchrotron).   Tandem van de Graaffs cannot reach the necessary current and are not considered.

\subsection{RFQ with Separated Sector Cyclotrons Running at 60 to 70~MeV}
One of the limits for achieving high power  in cyclotrons is the
amount of beam that can be captured at low energies for acceleration.
Conventional axial injection with a spiral inflector is subject to
current limits from space-charge blowup, and emittance dilution
because of phase-space coupling as the beam bends from the vertical
direction to the horizontal cyclotron plane.  These current limits have
been thoroughly explored with proton and H$^-$ beams injected into
high-current cyclotrons used for isotope production; the typical
limits (for $Q/A = 1$) for beams injected at 30~kV are around 2~mA.  Using perveance (Eq.~\ref{perveance})
to describe this,  the higher the value of $K$,  the more sensitive
a beam is to space-charge effects.  Taking this value of $K$
for protons (and 2~mA current) and scaling this for H$_2^+$ (with Q/A = 0.5),
we see that an electrical current of 5~mA (10~mA of protons)
will have the same $K$ value if the injection energy is raised to 35~keV/amu (terminal potential increased to 70~kV).  
This is the heart of the argument that 
H$_2^+$ beams will have substantially higher currents.
An experiment is under construction at Best Cyclotron Systems in
Vancouver~\cite{BESTtest} 
to verify this scaling (i.e. proof of the feasibility of high intensity H$_2^+$ injection using a spiral inflector). 
At the same time, we will further refine our numerical models for precise central region designs and parameter prediction for the IsoDAR cyclotron.  

The simplest alternative approach is reducing current and
increasing the injection energy.  
Eq.~\ref{perveance} shows that if the injection energy is increased
from 35 to 800~keV/amu, $K$ drops to only 1\% of its original
value, and injection space-charge issues become insignificant.  
This voltage would be too high to use a spiral inflector, 
but injection at higher energies is certainly feasible into a
separated-sector cyclotron.   The PSI Injector~\cite{PSI-II} is a good example: 12~mA of protons at 
800~keV are injected into a separated sector cyclotron and
$\sim$ 3~mA are accelerated up to 72~MeV.    The controlled losses (collimation) are at the first turns which do not produce high activation.  
Limits at PSI are determined by beam loss at the highest energies in the ring cyclotron and 
not by the performance of the injector cyclotron.

While PSI uses a Cockcroft-Walton platform to produce the energy of
800~keV, less expensive alternatives are possible today using RFQ
accelerators.  Such compact LINACs can be built now with parameters
quite suitable for injection into a cyclotron, namely CW operation 
(100\% duty factor), with very low energy spread, and with
an RF frequency that matches the optimum value for the cyclotron
itself.  In fact, the Indiana Cyclotron is injected by such an
RFQ, manufactured by AccSys Inc~\cite{AccSys}, an accelerator 
company located in Pleasanton, CA.  

A preliminary look at the IsoDAR configuration using an RFQ injecting a 
4-sector ring cyclotron indicates costs could increase by 40\% or
more from the baseline. With this said, the RFQ-injection alternative does provide a path forward in case the BEST Cyclotron Systems tests show an issue with the spiral inflector
injection.

\subsection{LINACs With Low Energy (30~MeV) and High Current (40~mA) \label{linear}}

Proton LINACs can provide the correct combination of beam energy and current for reaching the necessary $\bar \nu_e$ flux.   These machines are long
chains of accelerating RF cavities, where each cavity increases the beam
energy by as much as a few~MeV.
To increase electrical efficiency, modern high current LINACs are superconducting and therefore markedly decrease wall current losses, and thereby increase the
electrical efficiency.

As compared to cyclotrons, LINACs can provide much higher current with the same level of losses.  This is because strong
transverse-focusing can be applied using quadrupole magnets
placed throughout the line. Also, relatively strong longitudinal focusing
is produced by the high frequency of the RF cavities.   This can be
compared to the cyclotron, where both transverse and longitudinal focusing are
weak and therefore space charge is challenging to control at injection.
The result is that the limit of current in the LINAC will be largely
set by the output of the ion source.    In this discussion, we assume
that 40~mA can be accelerated.   Using Table~\ref{pnplot}, this sets
our LINAC energy at 30~MeV.

The size of a LINAC is driven by the fact that the beam passes through the line only once.
This necessarily requires a long machine, with many cavities, and high power needs are inevitable for both the RF and the cryogenic systems (for superconducting machines).  The length
is a function of beam energy and the accelerating gradient.   The
superconducting cavities offer higher gradients, reducing the length
by a factor of three.  Using the ESS as our example, the high current LINAC 
length is 27~m for 50~MeV in energy~\cite{Tord}.

A study by the IAEA~\cite{IAEA-TECDOC-1439} considered
a $d$-beam at $\sim$25~MeV, accelerated by a LINAC,
on $^7$Li as a neutron generator. The quoted cost for such a machine is ``$>$\$50M for 100~kW of power on target".  The most relevant high power, low energy ion accelerator being designed is the International Fusion Materials Irradiation Facility (IFMIF). IFMIF draws heavily from the experience at the LEDA at Los Alamos. The cost-scaling from the IFMIF design work is $>$\$100 M. As a reference point for the cost of RF-power and cryogenics, one can consider the costs for the Jefferson Lab CEBAF Upgrade, which was \$5M for the complex of 13~kW  RF-amplifiers, which would be needed for IsoDAR~\cite{Rimmer}.   This leads
to the same conclusion, namely that this solution will cost far more than \$50M and is therefore not cost-effective.

\subsection{FFAGs}
We have also considered a FFAG accelerator design. Such a machine would resemble a sector cyclotron, the difference being that the magnetic field varies radially as well as azimuthally. In some sectors the field increases with radius; in other sectors it decreases.  This alternating gradient produces strong focussing, so that the bunches in such machines are better contained than they are in a cyclotron, and losses are lower. If the scaling requirement --that the optics of the beam is independent of energy-- is relaxed, then there is considerable freedom in the choice of magnetic field configuration. If this field configuration is chosen carefully the design can also be isochronous, so the machine can operate continuously at a fixed RF frequency. The relaxation of scaling means that the particle bunch will pass through tune resonances during its acceleration. However, if the acceleration is fast then this need not be fatal, as has been shown by the success of the EMMA prototype~\cite{machida, garland}. This combination of focussing and CW operation, the best features of the synchrotron and the cyclotron, make the FFAG a very appealing choice for IsoDAR.

However, experience with FFAG machines is still limited. EMMA is an electron machine, and does not experience the problems of varying velocity, or space charge, that is expected with IsoDAR. The proton machines constructed at KURRI~\cite{tanigaki} have operated successfully in the energy range we require, but have only achieved currents in the nA-range. An FFAG for IsoDAR would involve a new design and a considerable R\&D program. We will continue to investigate the possibility of an FFAG - it may be that other applications and industries will be interested and drive these developments. However, the FFAG concept is still insufficiently proven for us to adopt it in preference over an established cyclotron design.

\subsection{Rapid Cycling Synchrotrons}

\begin{figure}[t]
\begin{center}
{\includegraphics[angle=-0, width=.95\linewidth]{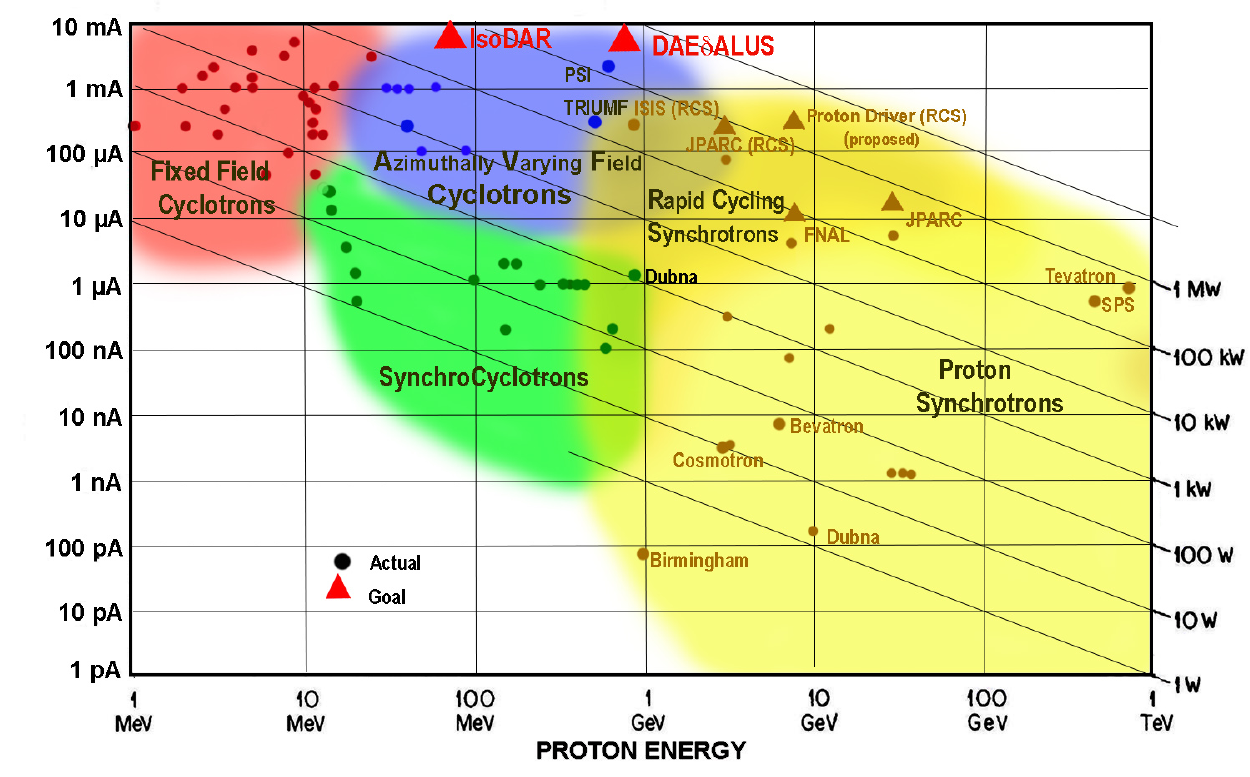}}
\end{center}
\vspace{-.2in}
\caption{Current versus proton energy for various existing machines. Various types of cyclotrons are shown; FF is the Fixed Field or Classical Cyclotron;
FM is the Frequency Modulation (Synchro-) Cyclotron; and 
AVF is the Azimuthal Varying Field Cyclotron.  One can see that 
IsoDAR is far from the space populated by Rapid Cycling Synchrotons.
This is an updated version of Fig. 1 from Ref~\cite{catalogue}.
\label{RCS} }
\end{figure}

Fig.~\ref{RCS} shows the context in which cyclotrons and rapid cycling synchrotrons are employed.    The machines are mapped in energy-current-beam power parameter space. The upper left corner
(in red) is the domain of fixed B-field cyclotrons. In the classic
Lawrence design the uppermost energy achievable is between 10 to 20~MeV depending on the RF-voltage. At higher energies the beam loses
synchronism with the RF. Synchronism can be restored by varying the
frequency of the RF (green area). The energy reach up to 1~GeV comes at the expense of much reduced beam
current as only a single bunch can be in the machine at any one time.
The upper energy is limited by the declining extraction efficiency as
the beam orbits get closer together. At 1~GeV, the extraction
efficiency of the Dubna machine has fallen to 30\%.  Much higher
energies can be achieved in proton synchrotrons.

The highest energy extracted beam (800~GeV) was possible with the
Tevatron.  The beam current in these machines is limited by the
Laslett incoherent tune shift at injection that is in turn set by the
beam emittance and injection energy.  The Tevatron had the capability
of delivering an average current of up to 500~$\mu$A in pulses at 1~Hz. Lower energy,  rapid cycling synchrotrons have been proposed and
built with pulsed beams at rates approaching 100~Hz. The maximum
current, again limited by the incoherent tune shift, is $\sim$300~$\mu$A.

Achieving very high power, especially at low proton energy, requires a
machine capable of accelerating 1 to 10~mA.  This regime is the
province of cyclotrons with a radially and azimuthally varying B-field that provides stronger edge focusing of the beam and assures
synchronism with a fixed RF-source over the entire acceleration cycle.
Such machines with both resistive and superconducting magnet coils are
mainstays of the research and medical industries. Extraction
efficiencies, even at hundreds of~MeV, exceed 99.9\%. Machine
reliability and availability is also very high ($\sim$95\%).  The
proposed DAE$\delta$ALUS and IsoDAR accelerators are of this variety and
build on decades of experience with this proven accelerator technology.

\section{Radically Different Designs \label{radicalalt}}

The previous discussion centered on maintaining the basic IsoDAR design while replacing individual elements. In this section, we
consider radically different designs.

\subsection{Why Not Use the $\beta$-beam Production Design?}

Designers of $\beta$-beam technology have studied the production of $^8$Li using a 25~MeV beam~\cite{Lindroos}. Perhaps IsoDAR could use the reaction $^7$Li$(d,p)^8$Li 
to produce the necessary flux at 25~MeV?   In this
section,  we explore this possibility and ultimately conclude that it is not feasible.

The $\beta$ beam is produced by impinging a $^7$Li beam on a gas deuterium
target to produce $^8$Li, for subsequent acceleration.  
This type of reaction, where the beam is more massive than the target,
is referred to as ``reverse kinematics." The 25~MeV energy is just
above Coulomb barrier in the center of mass system for this asymmetric
target-beam species. Beta beams obviously need as much motion {\it in} 
the center of mass frame as possible, for boosting acceleration and
kinematic focusing to form a beam; however, 
this is irrelevant and in fact not desirable for the case of IsoDAR. 
A gas target is used so that the relatively fast outgoing $^8$Li isotope 
can be extracted for acceleration.  The $^7$Li is stored in a recirculating ring in order to increase the rate~\cite{Rubbia}.   

There are two issues which cause us to reject this base design.  Even
with the recirculating design, the maximum production will be about
$3\times 10^{21}$~ions per year~\cite{Lindroos}, which is
10 times lower than the baseline IsoDAR flux.    Therefore, it is a high-risk assumption
that solutions could be found to increase rates by an order of
magnitude.  Further, the system requires a two accelerator chain,
first to reach 25~MeV, and then to recirculate the beam.
This will increase cost, add complexity, and enlarge the necessary
footprint of the design and leads us to the conclusion that adapting this
recirculating-beam-gas-target design for IsoDAR would be much more expensive than the baseline cyclotron system proposed.

On this basis, we reject the use of the $\beta$-beam design.  However,
we can now investigate designs ``inspired'' by the planned $\beta$-beam 
production.

\subsubsection{$^7$Li Beam on a Liquid Deuterium Target}

A small LINAC
could be used to accelerate $^7$Li up to $\sim$25~MeV to impinge on a liquid
deuterium target. However, we conclude that this is not a cost-effective option as LINACs are quite expensive 
(discussed in Sec.~\ref{linear}) and because maintaining a cryogenic liquid target
in a system that absorbs nearly 1~MW of beam power is challenging.

Why not use a tandem accelerator since this would be less expensive than an RF LINAC?
A tandem is a high impedance device which
therefore delivers low beam currents continuously.
A high current tandem might deliver 10 $\mu$A.      The advantage of
the tandem is extreme voltage stability (0.5\%)  that can be important
for low energy nuclear physics, when one wants to sit on a resonance.
However, such an accelerator does not fulfill the requirements of IsoDAR.

\subsubsection{Deuteron Beam on a $^7$Li Target}

Classic measurements of the reaction $d +^7$Li$\rightarrow p+^8$Li have been
done with few~MeV beams on solid lithium
targets~\cite{FilliponePhysRevC25p2174}.     A low energy beam is
appropriate for exploiting resonances of hundreds of millibarns in the deuteron energy region of 3-4~MeV.    However, low energy deuterons range out very
quickly, penetrating $<0.5$~mm into target.   Energy loss moves
the deuteron through the resonance quickly, so that $<0.1\%$ of the
beam particles interact to produce $^8$Li.   In the designs we have
explored, the rates of $^8$Li production are two orders of magnitude lower than the baseline IsoDAR design.

\subsection{Use of Existing or Planned Accelerators and a New Detector}

The last alternative to consider is whether it is more cost-effective to
build a new detector at an existing accelerator facility or build a new accelerator at an existing large scintillator-based detector as is proposed for IsoDAR.

A cyclotron and a 1~kton ``KamLAND-like'' detector,
such as the one proposed for OscSNS~\cite{OscSNS}, 
are approximately the same cost.  This can be seen by comparing 
the cost  estimates in  Ref.~\cite{OscSNS} and
Ref.~\cite{cost-cycl-2005}.     Thus, the issue is not in the cost of the 
equipment itself, it is in the cost of the implementation.

Oil-based detectors at the surface require small-duty
factor beams because of backgrounds produced by spallation from 
cosmic-ray muons and from decays of stopping muons.  
The IsoDAR source cannot make use of a small duty-factor beam
because the half-life of $^8$Li is 841~ms.   Therefore, the flux will be
continuous, regardless of beam structure, unless the beam bunch spacing
is significantly more than 1~s.     There are no existing facilities
that offer 600~kW of beam power at $\sim 60$~MeV with pulsed 
spacing of $\gg 1$~s. With the 6\% duty factor for LSND, achieving 600~kW
at 60~MeV would require 160~mA of current on target.

It is for this reason that scintillator detectors  
used with continuous beams are built
underground.   Recent examples are the modern
reactor experiments, which have very similar rates of
IBD interactions in their far-detector halls as IsoDAR.     The Double Chooz far
hall is at 300~mwe depth~\cite{DC}, the RENO far hall is at 450~mwe~\cite{RENO} and the Daya Bay EH3 far hall is at  860~mwe~\cite{DB}.   These choices set the scale 
for the range of acceptable depth for a detector accomplishing IsoDAR-like
physics.

A new detector that meets IsoDAR's physics goals 
would need to be installed hundreds of feet underground.
There is no existing facility where this 
civil construction would be cheap.   One can consider the cost
estimates for the near hall of LBNE, which was proposed to be 185~ft
below grade, to see that this hall would be a very expensive project.
Along with the experimental hall, a new beamline would be required to bring the beam to the detector as well. For these reasons, we do not consider deploying a new detector at an existing accelerator as cost-effective.

\section{Discussion and Conclusions \label{conclude}}

We now compare the IsoDAR design to the most often suggested
alternative proposals: the RFQ/Separated Sector Cyclotron
design;  a LINAC design;  a modified $\beta$-beam design using a low
energy deuteron beam on a $^7$Li target; and 
a new detector, matching KamLAND specifications,  built under 300~m.w.e shielding at an existing accelerator laboratory.   

\begin{figure}[t]
\begin{center}
{\includegraphics[angle=-0, width=1.\linewidth]{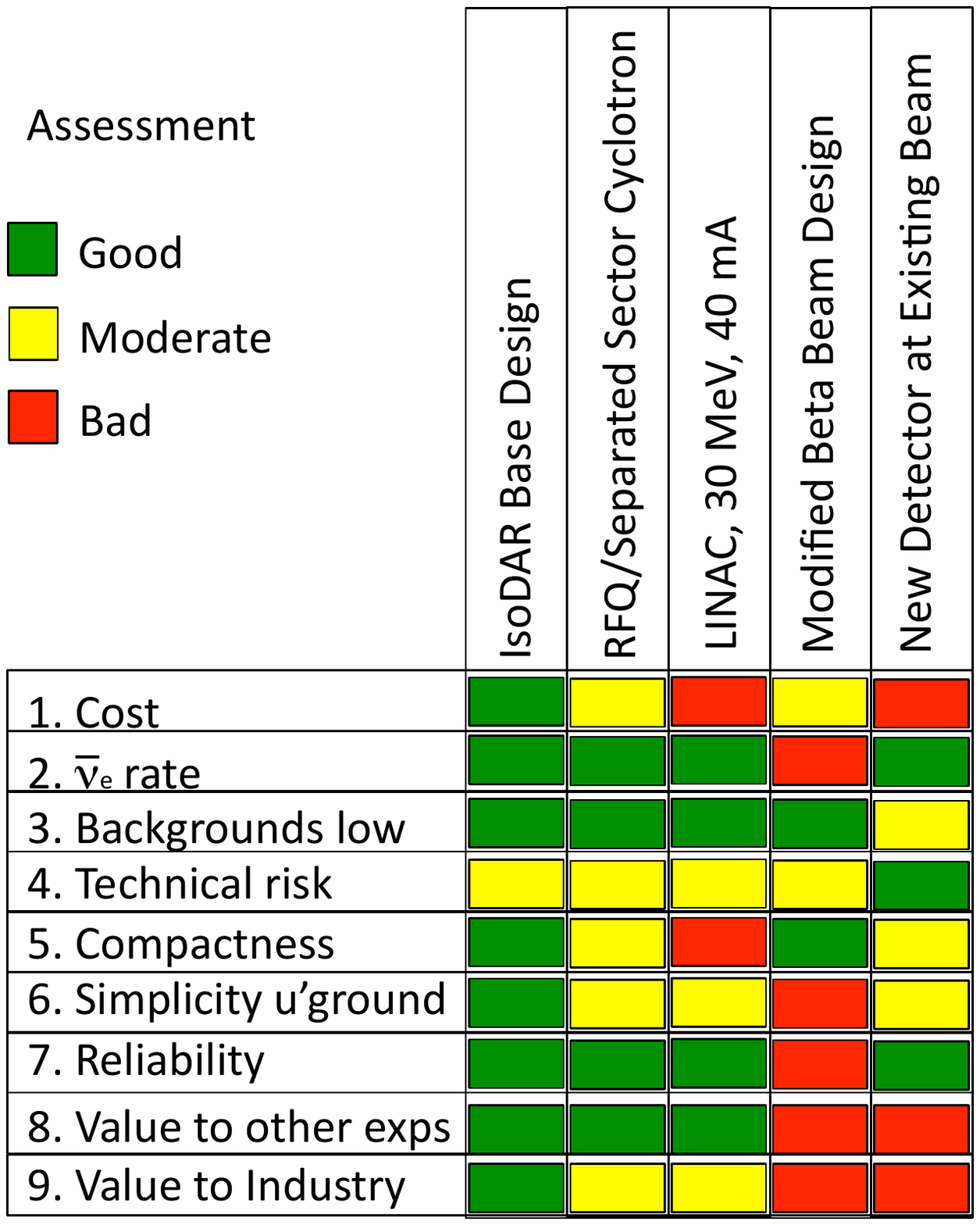}}
\end{center}
\vspace{-0.7in}
\caption{Comparison of IsoDAR to alternative designs.  See text for
  explanation. \label{final}}
\end{figure}

As shown in Fig.~\ref{final},
we roughly classify
the designs considered as ``good" (green), ``moderate" (yellow), or ``bad" (red) for each criterion. The meaning of the grade for each criterion is described below:
\begin{itemize}
\item {\it Cost:}~~Good: $\sim\$30$M, Moderate: $\sim\$50$M,
  Bad: $\sim\$100$M or higher.   The cost estimates can be considered very rough.
  Plausibility of the IsoDAR
  estimate can be cross checked against the D.O.E. cost study of a 70~MeV, 1~MW cyclotron~\cite{cost-cycl-2005}.    In the case of a new
  experiment, the cost includes a new beamline and an underground
  site, as well as the KamLAND-like detector.
\item  {\it $\bar \nu_e$ Rate in 5 years:}~~Good: $\gtrsim 1\times10^{23}$;
  Moderate: $\sim 5\times10^{22}$;  Bad: $< 1\times10^{22}$.
\item  {\it Backgrounds:}~~For IsoDAR and the first four
  comparisons, Good: $<1\%$ $\nu_e$; Moderate: $<5\%$ $\nu_e$; Bad: $>5\%$
  $\nu_e$  with endpoint $> 3$~MeV.   In the case of a new detector,
  the background at 300~mwe will overwhelm the singles signal. However, the background should be adequate for the IBD-based measurements, and is therefore marked as moderate.
\item{\it Low technical risk:}~~Good:  Very little R\&D required, uses
  proven technology;
  Moderate:  Modest R\&D required, uses cutting-edge technology; Bad:
  Significant R\&D needed, uses unproven technology.   IsoDAR makes
  use of the new VIS source, and is therefore marked as moderate.
\item {\it Compactness of both 
accelerator and source:}~~Good:  Expect very little underground
excavation; Moderate: requires new rooms of conventional size;  Bad:
requires major construction.   
For the new beamline, new detector alternative, we
assume the laboratory will have sufficient  space on site.
\item {\it Simplicity of underground construction and operation:}~~Good:  Excellent modularity;  Moderate:  must plan around some large
pieces; Bad: many large pieces that do not fit through tunnels.   
\item{\it Reliability:}~~Good:  95\% uptime typical, so unlikely to
  be an issue;  Moderate: 90\%
  uptime typical, so could be the limiting factor;     Bad: $<$90\%
  uptime, so the limiting factor.   All of the designs are expected to
  be highly reliable except for the modified $\beta$-beam design which
  is likely to have significant technical difficulties.
\item {\it Value
to future physics programs:}~~Good:  multiple examples of
applications in physics;  Moderate: one other example;
Bad: no examples.   In the case of IsoDAR, these include application
of the technology to DAE$\delta$ALUS and to rare isotope production
facilities such as Legnaro, Holifield, and the 70~MeV cyclotron in Nantes.
\item {\it Value of this development to industry:}~~Good:  multiple examples of
interested industries;  Moderate: one other example;
Bad: no examples.   In the case of IsoDAR, the IBA and BEST Cyclotron Systems companies
have both demonstrated interest in the design.
\end{itemize}

Based on this study, we conclude that the IsoDAR base design is the
best technology choice for the planned physics application.



%
%
\end{document}